\documentclass[journal,draftcls,onecolumn,12pt,twoside]{IEEEtran}
\usepackage{textcomp}
\usepackage{stfloats}
\usepackage{url}
\usepackage{verbatim}
\usepackage[T1]{fontenc}
\usepackage{ifpdf}

\usepackage{cite}

\usepackage[dvips]{graphicx}
\graphicspath{{../}}
\usepackage{graphicx}
\DeclareGraphicsExtensions{.pdf,.jpeg,.png,.eps}

\usepackage{amsthm}
\usepackage{amsmath}
\usepackage{amsfonts}
\usepackage{amssymb}
\usepackage{mathrsfs}
\usepackage[cmintegrals]{newtxmath}
\usepackage{bm}

\newtheorem{lemma}{\textbf{Lemma}}
\newtheorem{theorem}{\textbf{Theorem}}

\newtheorem{definition}{\textbf{Definition}}
\newtheorem{example}{\textbf{Example}}
\newtheorem{remark}{Remark}
\newtheorem{corollary}{\textbf{Corollary}}

\usepackage[ruled,linesnumbered]{algorithm2e}
\usepackage{array}
\usepackage{multirow}
\usepackage[table,xcdraw]{xcolor}
\usepackage{booktabs,threeparttable}
\usepackage[caption=false,font=normalsize,labelfont=sf,textfont=sf]{subfig}

\begin{document}
\title{Capacity-Achieving BBT Polar Codes\\
with Interleaver-Assisted BP Decoding}
\author{Xinyuanmeng~Yao
and Xiao~Ma~\IEEEmembership{Member,~IEEE}
\thanks{

Xinyuanmeng Yao is with the School of Cyber Science and Engineering and also with Ningbo Key Laboratory of Information Technology Application   Innovation and Security, Ningbo University of Technology, Ningbo 315211, China. She is currently a visiting scholar with the Department of Electrical and Electronic Engineering, Imperial College London, London SW7 2AZ, U.K. (e-mail: yaoxym@nbut.edu.cn).

Xiao Ma is with the School of Computer Science and Engineering, and also with Guangdong Province Key Laboratory of Information Security Technology, Sun Yat-sen University, Guangzhou 510006, China (e-mail: maxiao@mail.sysu.edu.cn). 

The corresponding author is Xiao Ma.}}

\maketitle

\begin{abstract}
In this paper, we introduce a binary balanced tree (BBT) channel transformation that extends Arıkan's channel transformation to arbitrary block lengths. We prove that the proposed transformation induces channel polarization, thereby establishing that BBT polar codes achieve the capacity of binary-input memoryless symmetric~(BMS) channels. To characterize the finite-length performance of BBT polar codes, we further develop an efficient method for estimating the weight spectrum by exploiting the hierarchical tree structure, and derive analytical upper and lower bounds on the frame error rate~(FER) under maximum-likelihood~(ML) decoding.

For practical low-latency implementations, we propose interleaved BBT~(IBBT) polar codes together with a belief-propagation (BP) decoding algorithm. Specifically, based on the normal-graph representation of BBT polar codes, interleavers are introduced between adjacent layers to modify the message-passing schedule. In addition, we propose to perform BP decoding on an IBBT sub-normal graph and replace partial BP processing modules with a posteriori probability~(APP) calculation modules, thereby reducing the number of message-passing steps required per iteration. Numerical results demonstrate that the proposed interleaving strategy improves decoding convergence, while the sub-normal-graph-based BP decoding algorithm significantly reduces decoding latency while maintaining comparable error-rate performance.
\end{abstract}


\section{Introduction}\label{section1}
Polar codes, introduced by Arıkan in~\cite{Arikan2009Polarization}, are a class of provably capacity-achieving codes for binary-input memoryless symmetric~(BMS) channels. Their fundamental principle is channel polarization, whereby several independent and identically distributed~(IID) BMS channels are transformed into synthetic channels whose reliabilities asymptotically approach either a noiseless channel or a completely noisy channel. As a result, reliable communication can be achieved by transmitting information bits through the highly reliable synthetic channels while freezing the others. Owing to their solid theoretical foundations and excellent finite-length performance under cyclic redundancy check~(CRC)-aided successive cancellation list~(SCL) decoding~\cite{Niu2012CRCAided,Tal2015ListDecoding}, polar codes have been adopted for control-channel coding in the 5G new radio~(NR) standard~\cite{3GPP2018NR38212}.

The original polar codes rely on the polarization effect induced by the Kronecker powers of the binary kernel matrix {\tiny $\left(\begin{array}{ll}1 & 0 \\ 1 & 1\end{array}\right)$}. Consequently, the block lengths are restricted to powers of two. This structural constraint is not that convenient for some applications in modern communication systems, which often require flexible block lengths and fine-grained rate matching. To address this limitation, a variety of approaches have been proposed to construct length-flexible polar codes. Two representatives are punctured and shortened~(P/S) polar codes~\cite{Eslami2011PracticalPolar,Niu2013RateCompatible,Wang2014Puncturing,Miloslavskaya2015Shortened,Shin2013LengthCompatible,Jang2019BinaryDominationRateMatching,Zhao2021LowRatePuncturing,Han2022RateCompatiblePunctured}, which are obtained by removing selected coded bits from a mother polar code. Although P/S polar codes can achieve good performance, decoding is still carried out with respect to their mother codes, which typically increases decoding latency. When the target block length is slightly larger than a power of two or the code rate is low, extended polar codes~\cite{Chen2013PolarHARQ,Saber2015PolarIRHARQ,Ma2017IRHARQPolar,Zhao2018AdaptiveIRHARQ,Jang2020SimplexKernelPolar}, which append additional bits to a shorter polar code, may provide a more favorable performance–latency trade-off. In the 5G standard, a carefully designed rate-matching scheme combining puncturing, shortening, and repetition is adopted to enable polar codes compatible with arbitrary block lengths~\cite{3GPP2018NR38212}. Beyond these three techniques, several alternative constructions have been proposed to enable length flexibility in polar codes. For example, multi-kernel polar codes~\cite{Bioglio2020MultiKernel} combine kernels of different sizes, chained polar subcodes~\cite{Trifonov2018RandomizedChainedPolar} exploit the decomposition of integers into sums of powers of two, and asymmetric polar codes~\cite{Cavatassi2019AsymmetricPolarConstruction} interconnect polar codes of unequal lengths. Despite these advances, the design of polar-code structures that simultaneously accommodate arbitrary block lengths, provide rigorous theoretical guarantees, and deliver strong practical performance remains an open and active research problem.

In our previous work~\cite{Yao2024BalancedTree}, we proposed a class of length-flexible polar codes based on a balanced binary tree~(BBT) structure. Specifically, by constructing a BBT and defining a length-adaptive $(U+V,V)$ relationship between each parent node and its two child nodes, BBT polar codes can be obtained. The resulting codes preserve the recursive encoding and decoding structure of classical polar codes while offering greater flexibility in code length. Numerical results show that BBT polar codes provide potential advantages in both error-correction performance and decoding latency compared with other existing length-flexible polar codes.

This paper further develops the theoretical foundations and the practical implementations of BBT polar codes. Our main contributions are summarized as follows.
\begin{enumerate}
\item To establish the capacity-achieving property of BBT polar codes, we introduce a BBT channel transformation that generalizes the classical polar transformation. Under this transformation, an arbitrary number of IID BMS channels are recursively combined to generate synthetic channels. Building upon the proof technique in~\cite{Alsan2014PolarizationProof}, we show that the fraction of reliable synthetic channels converges to the channel capacity as the code length grows. Consequently, BBT polar codes under successive cancellation~(SC) decoding achieve the capacity of BMS channels. 

\item Leveraging the hierarchical subcode decomposition inherent in the BBT structure, we propose a weight enumerating function~(WEF) estimation method tailored to BBT polar codes. Based on this method, we further develop a fast minimum Hamming weight~(MHW) estimation algorithm. By combining the estimated WEF with existing maximum-likelihood~(ML) performance bounding techniques, we derive analytical upper and lower bounds on the frame error rate~(FER) of BBT polar codes under ML decoding.

\item To enable efficient low-latency decoding, we investigate an interleaver-assisted belief-propagation~(BP) decoding algorithm for BBT polar codes. Motivated by the observation that interleaving can improve the convergence behavior of iterative decoding algorithms, we introduce interleavers between adjacent layers of the normal graph representation of BBT polar codes. The resulting interleaved BBT~(IBBT) polar codes are analogous to interleaved polar codes~\cite{Chiu2020IPolar}. Furthermore, we propose to perform BP decoding on an IBBT sub-normal graph and replace partial BP processing modules at the bottom layers with a posteriori probability~(APP) calculation modules, thereby reducing the number of message-passing steps required per iteration.
\end{enumerate}

The remainder of this paper is organized as follows. Section~II reviews the BBT channel transformation and the BBT polar codes. Section~III presents the BBT channel polarization theorem. Section~IV introduces the weight spectrum estimation method and derives ML performance bounds. Section~V presents IBBT polar codes with BP decoding. Finally, Section~VI concludes the paper.

\section{Preliminaries}\label{section2}
In this section, we first introduce the concept of a BBT, which serves as the basis for a generalized channel transformation, referred to as the BBT channel transformation. This transformation can map an arbitrary number of IID BMS channels into different synthetic channels. We then present two equivalent representations of the BBT channel transformation: a tree graph and a normal graph. Based on the tree-graph representation, we briefly review the BBT polar coding scheme proposed in~\cite{Yao2024BalancedTree}. The normal-graph representation will be used in parts of the remainder of this paper.
\subsection{BBT Structure}
A BBT is a binary tree in which, for every internal node, the heights of its left and right subtrees differ by at most one~\cite{Tenenbaum1990DataStructures}. In this work, we adopt the BBT structure defined as follows.

\begin{figure*}[t]
\centering
\begin{minipage}[t]{0.38\textwidth}
\vskip-5cm
\centering
\includegraphics[width=0.7\textwidth]{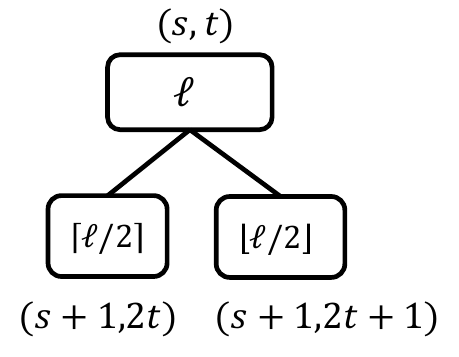}
\vskip1.5cm
\caption{The parent-child relationship in a BBT.}
\label{NodeRelationship}
\end{minipage}
\hfill
\begin{minipage}[t]{0.58\textwidth}
\centering
\includegraphics[width=\textwidth]{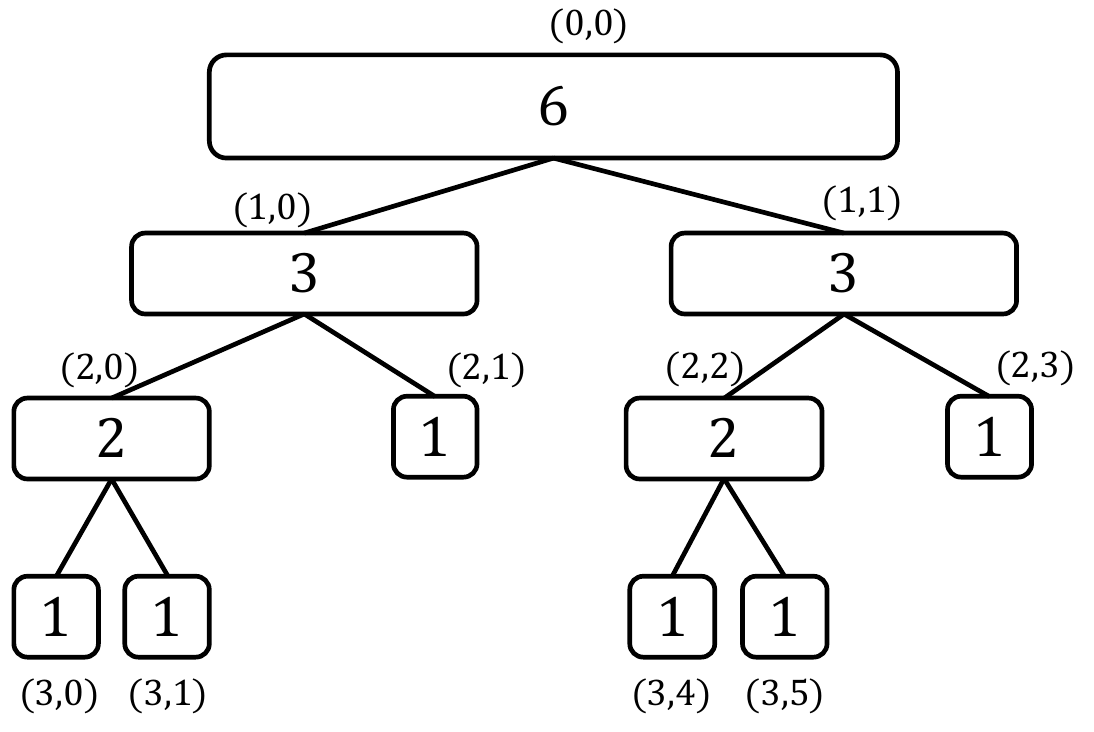}
\caption{The BBT structure for $N = 6$.}
\label{BBTStructureForN6}
\end{minipage}
\end{figure*}

In the BBT, the root node is located at level~$0$ and is associated with length~$N$. Each internal node with length $\ell \leq N$ (not necessarily a power of two) has two children: a left child of length~$\lceil \ell/2 \rceil$ and a right child of length~$\lfloor \ell/2 \rfloor$, where $\lceil \cdot \rceil$ and $\lfloor \cdot \rfloor$ denote the ceiling and floor functions, respectively. To index the nodes in the tree, each node is assigned a pair~$(s,t)$. The root node is indexed by~$(0,0)$. For a parent node indexed by~$(s,t)$, its left child is indexed by~$(s+1,2t)$ and its right child by~$(s+1,2t+1)$. The parent–child relationship is illustrated in Fig.~\ref{NodeRelationship}.

According to the above definition, for any code length~$N$, the corresponding BBT can be constructed recursively starting from the root by splitting each node with $\ell \geq 2$ into two children. The resulting BBT has $n+1$ levels and exactly $N$ leaf nodes, where $n \triangleq \lceil \log_2 N \rceil$. An example of the BBT for $N=6$ is shown in Fig.~\ref{BBTStructureForN6}.
\subsection{BBT Channel Transformation}
Let $W_{0}$ and $W_{1}$ be two BMS channels with input alphabet $\mathcal{X}=\{0,1\}$, output alphabet $\mathcal{Y}$, and transition probabilities $W_{0}(y|x)$ and $W_{1}(y|x)$, respectively, where $x\in\mathcal{X}$ and $y\in\mathcal{Y}$. The polar transform combines the two channels into a pair of synthetic channels, an ``upper'' synthetic channel $W_{0}\ast W_{1}$ defined as
\begin{align}
W_{0}\ast W_{1}(y_{0},y_{1}\mid x_{0})
= \sum_{x_{1}\in\{0,1\}} \frac{1}{2}
W_{0}(y_{0}\mid x_{0}\oplus x_{1})\, W_{1}(y_{1}\mid x_{1}),
\end{align}
and a ``lower'' synthetic channel $W_{0}\circ W_{1}$ defined as
\begin{align}
W_{0}\circ W_{1}(y_{0},y_{1},x_{0}\mid x_{1})
= \frac{1}{2}
W_{0}(y_{0}\mid x_{0}\oplus x_{1})\, W_{1}(y_{1}\mid x_{1}),
\end{align}
where $\oplus$ denotes addition modulo~2.

Based on the polar transform, a BBT channel transformation can be applied to an arbitrary number of IID BMS channels. To represent the BBT channel transformation, we adopt two equivalent graphical representations, the tree graph and the normal graph, which are described below.
\subsubsection{Tree-Graph Representation}
\begin{figure}[t]
\centering
\includegraphics[width=0.8\textwidth]{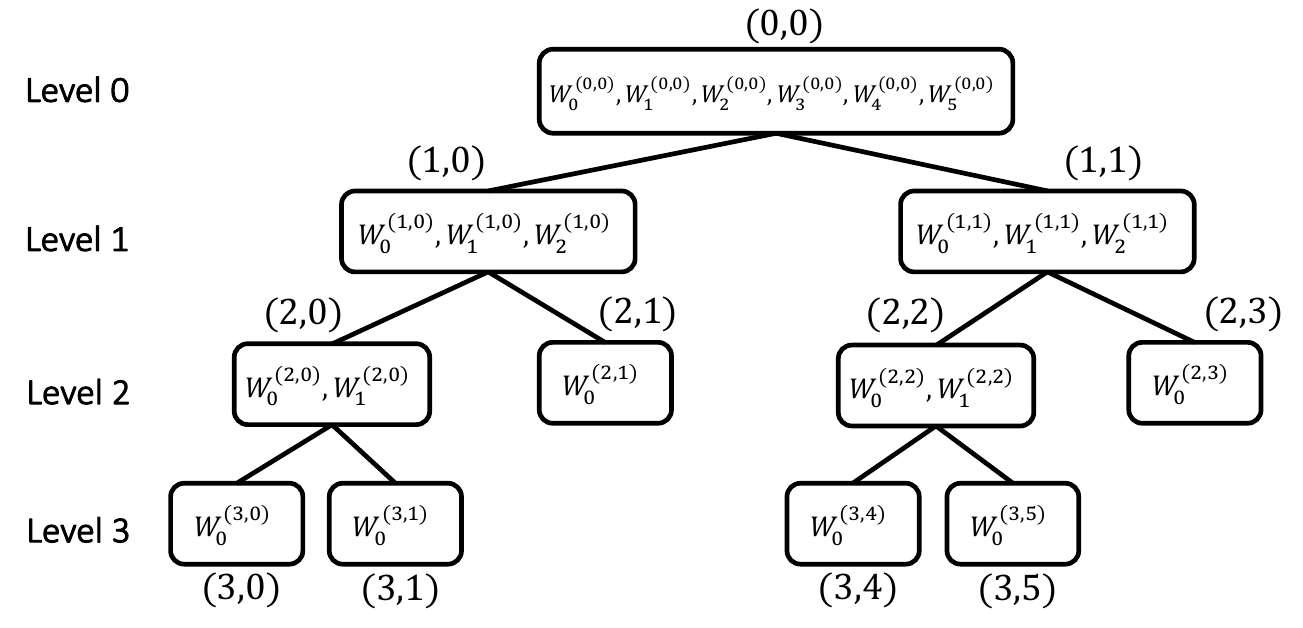}
\caption{Tree-graph representation of the BBT channel transformation for $N=6$.}
\label{BBTChannelTransformationForN6}
\end{figure}

The BBT channel transformation naturally admits a tree-graph representation induced by the underlying BBT structure for code length $N$. An example of this representation for $N=6$ is illustrated in Fig.~\ref{BBTChannelTransformationForN6}.

In the tree graph, each node $(s,t)$ with length $\ell$ is associated with a sequence of $\ell$ BMS channels. The root node $(0,0)$ is initialized with the channel sequence $(W^{(0,0)}_0,\ldots,W^{(0,0)}_{N-1})$, where $W^{(0,0)}_i = W$ for $0 \le i \le N-1$. For an internal node $(s,t)$ with the channel sequence $(W^{(s,t)}_0,\ldots,W^{(s,t)}_{\ell-1})$, the channel sequences of its two children are obtained by applying the basic channel operations ``$\ast$'' and ``$\circ$'' in a pairwise manner. Specifically, the channel sequence at the left child $(s+1,2t)$ is given by
\begin{align}
W^{(s+1,2t)}_{i}
= W^{(s,t)}_{i} \ast W^{(s,t)}_{\lceil \ell/2 \rceil + i},
\quad 0 \le i \le \lceil \ell/2 \rceil - 1,
\end{align}
when $\ell$ is even, and by
\begin{align}
W^{(s+1,2t)}_{i}
=
\begin{cases}
W^{(s,t)}_{i} \ast W^{(s,t)}_{\lceil \ell/2 \rceil + i},
& 0 \le i \le \lceil \ell/2 \rceil - 2, \\
W^{(s,t)}_i,
& i = \lceil \ell/2 \rceil - 1,
\end{cases}
\end{align}
when $\ell$ is odd. The channel sequence at the right child $(s+1,2t+1)$ is given by
\begin{align}
W^{(s+1,2t+1)}_{i}
=
W^{(s,t)}_{i} \circ W^{(s,t)}_{\lceil \ell/2 \rceil + i},
\quad
0 \le i \le \lfloor \ell/2 \rfloor - 1 .
\end{align}
Recursively applying this procedure from the root to the leaves yields $N$ synthetic channels corresponding to the $N$ leaf nodes of the tree.
\subsubsection{Normal-Graph Representation}
\begin{figure}[t]
\centering
\includegraphics[width=\textwidth]{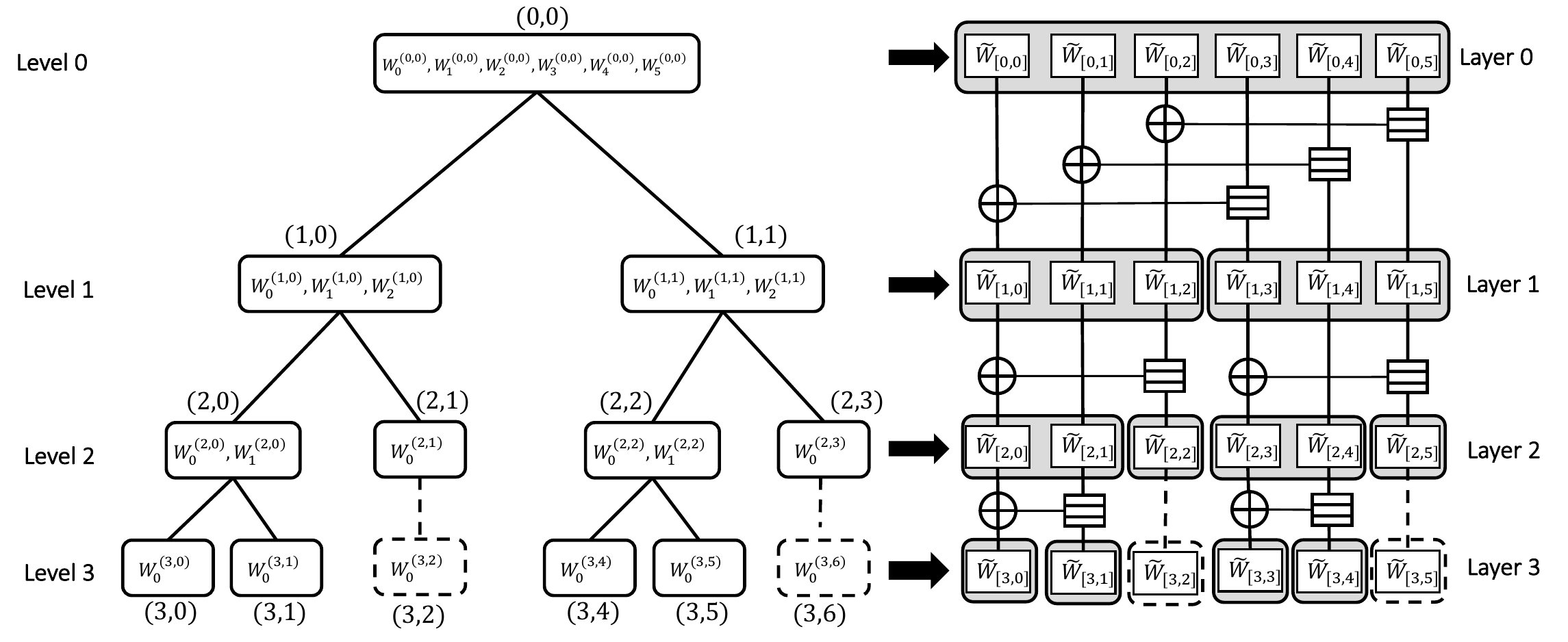}
\caption{Transformation from the tree graph to the normal graph for the BBT channel transformation with $N=6$.}
\label{TreeGraphAndNormalGraphForChannelTransformationForN6}
\end{figure}
The BBT channel transformation can also be represented equivalently by a normal graph obtained from the tree graph. The transformation from the tree graph to the normal graph for $N=6$ is illustrated in Fig.~\ref{TreeGraphAndNormalGraphForChannelTransformationForN6}, where the circle--angle--gray-square symbols in the normal graph indicate the corresponding tree nodes in the original tree graph.

To construct the normal graph, the tree graph is first slightly modified. For each leaf node at the $(n-1)$-th level (i.e., the penultimate level) with index $(n-1,t)$, a duplicate copy is introduced and treated as its child at the $n$-th level (i.e., the last level) with index $(n,2t)$, such that $W^{(n,2t)}_0 = W^{(n-1,t)}_0$. 

Based on this modification, we construct a normal graph consisting of $n+1$ layers, each containing $N$ variable nodes, where each node is associated with a channel. Specifically, the $j$-th variable node from left to right at the $i$-th layer from top to bottom is indexed by $[i,j]$\footnote{Throughout this paper, nodes in the normal graph are indexed using square brackets $[\cdot]$, while nodes in the tree graph are indexed using round brackets $(\cdot)$.}, and the corresponding channel is denoted by $\widetilde{W}_{[i,j]}$, for $0 \le i \le n$ and $0 \le j \le N-1$. Each channel $\widetilde{W}_{[i,j]}$ corresponds to the $j$-th channel at level $i$ of the tree graph; equivalently, it is the $q$-th component of the channel sequence associated with some node $(i,t)$.

In the resulting normal graph, each polar transform is represented by a pair of functional nodes: a sum node corresponding to the generation of the ``upper'' synthetic channel and an equality node corresponding to the generation of the ``lower'' synthetic channel.
\subsection{BBT Polar Codes with SC Decoding}
\subsubsection{Encoding}
\begin{figure}[t]
\centering
\includegraphics[width=0.5\textwidth]{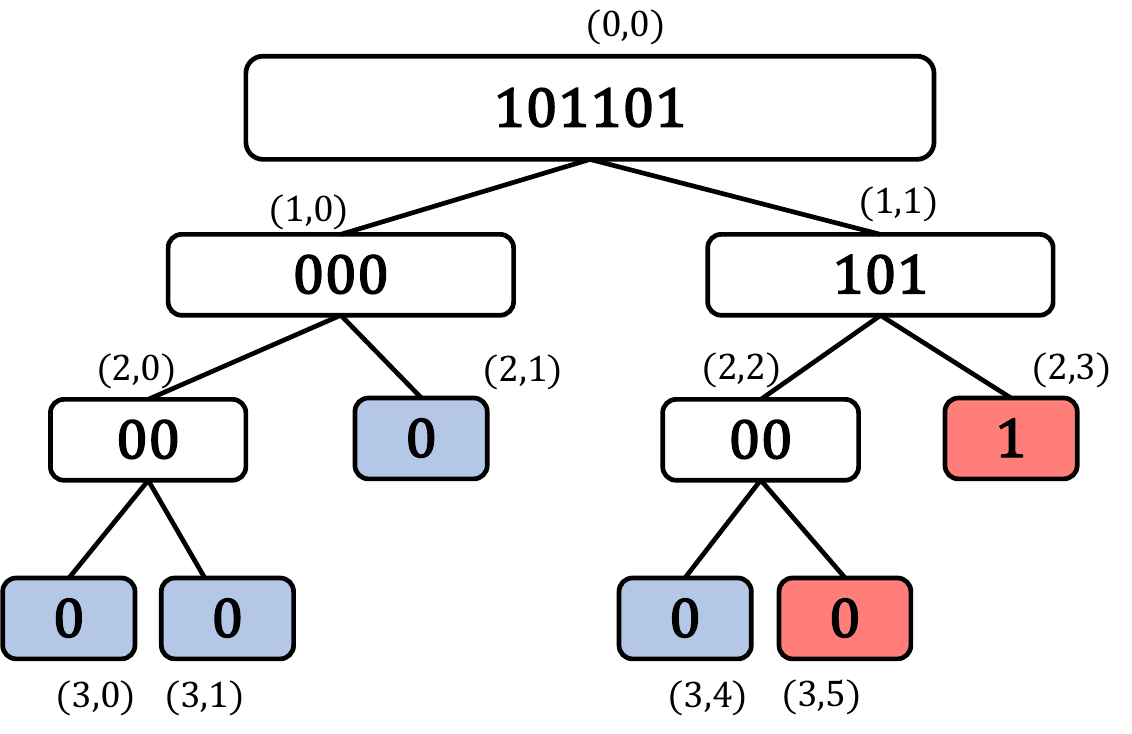}
\caption{The BBT polar encoding for $N = 6$ and $K=2$, where the blue leaf nodes are frozen and red leaf nodes are active.}
\label{BBTEncodingForN6}
\end{figure}
For encoding, we may also employ a tree graph derived from the BBT structure, in which each tree node of length~$\ell$ is associated with a code vector of length~$\ell$. In the tree graph for BBT polar encoding, a parent node $(s,t)$ of length~$\ell$ has two children, $(s+1,2t)$ and $(s+1,2t+1)$, with lengths $\lceil \ell/2 \rceil$ and $\lfloor \ell/2 \rfloor$, respectively. The parent code vector $\boldsymbol{v}^{(s,t)} \in \mathbb{F}_2^{\ell}$ and the corresponding child code vectors
$\boldsymbol{v}^{(s+1,2t)} \in \mathbb{F}_2^{\lceil \ell/2 \rceil}$ and
$\boldsymbol{v}^{(s+1,2t+1)} \in \mathbb{F}_2^{\lfloor \ell/2 \rfloor}$
are related by
\begin{equation}
\boldsymbol{v}^{(s,t)}
=
\Big(
\boldsymbol{v}^{(s+1,2t)} \boxplus \boldsymbol{v}^{(s+1,2t+1)},\,
\boldsymbol{v}^{(s+1,2t+1)}
\Big),
\end{equation}
where ``$\boxplus$'' denotes a length-dependent bit-wise addition modulo~2, defined as follows.

\begin{itemize}
\item If $\lceil \ell/2 \rceil = \lfloor \ell/2 \rfloor$, then
\begin{equation}
\boldsymbol{v}^{(s+1,2t)} \boxplus \boldsymbol{v}^{(s+1,2t+1)}
=
\big(
v^{(s+1,2t)}_0 \oplus v^{(s+1,2t+1)}_0,\,
\ldots,\,
v^{(s+1,2t)}_{\lceil \ell/2 \rceil - 1} \oplus v^{(s+1,2t+1)}_{\lceil \ell/2 \rceil - 1}
\big).
\end{equation}

\item If $\lceil \ell/2 \rceil = \lfloor \ell/2 \rfloor + 1$, then
\begin{equation}
\boldsymbol{v}^{(s+1,2t)} \boxplus \boldsymbol{v}^{(s+1,2t+1)}
=
\big(
v^{(s+1,2t)}_0 \oplus v^{(s+1,2t+1)}_0,\,
\ldots,\,
v^{(s+1,2t)}_{\lceil \ell/2 \rceil - 2} \oplus v^{(s+1,2t+1)}_{\lceil \ell/2 \rceil - 2},\,
v^{(s+1,2t)}_{\lceil \ell/2 \rceil - 1}
\big).
\end{equation}
\end{itemize}

Assume that the code length is $N$. To encode a data vector $\boldsymbol{u}\in\mathbb{F}_2^{K}$, the encoder assigns the data bits to $K$ selected active leaf nodes and sets all $N-K$ frozen leaf nodes to zero\footnote{The selection of active and frozen leaf nodes have been discussed in~\cite{Yao2024BalancedTree}.}. The code vectors of all internal nodes are then computed in a reverse level-order traversal of the tree, resulting in the transmitted codeword $\boldsymbol{c} = \boldsymbol{v}^{(0,0)}$ at the root. We provide an example in Fig.~\ref{BBTEncodingForN6}, where the data bits $\boldsymbol{u}=(0,1)$ is encoded into the codeword $\boldsymbol{c}=(1,0,1,1,0,1)$.

\subsubsection{BBT Polar Codes}
Based on the above construction, the BBT polar code is defined as follows. Index the leaf nodes from left to right by $\{0,1,\ldots,N-1\}$ and denote by $\mathcal{A}$ the information set and by $\mathcal{F}=\mathcal{A}^c$ the frozen set, where $|\mathcal{A}|=K$.

Let $\boldsymbol{x}\in\mathbb{F}_2^N$ be the vector formed from the leaves such that
$\boldsymbol{x}_{\mathcal{A}}=\boldsymbol{u}$ and
$\boldsymbol{x}_{\mathcal{F}}=\boldsymbol{0}$.
The resulting BBT polar code is given by
\begin{align}
\mathcal{C}_{\mathrm{BBT}}[N,K,\mathcal{A}]
=
\left\{
\boldsymbol{x}\boldsymbol{G}_{\rm BBT}
\;\middle|\;
\boldsymbol{x}_{\mathcal{A}}\in\mathbb{F}_2^{K},\;
\boldsymbol{x}_{\mathcal{F}}=\boldsymbol{0}
\right\},
\end{align}
where $\boldsymbol{G}_{\rm BBT}\in\mathbb{F}_2^{N\times N}$ is the invertible transformation matrix determined by the BBT structure of length $N$.

\subsubsection{SC Decoding}
For SC decoding, a tree graph derived from the BBT structure can similarly be employed by associating each node of length~$\ell$ with a log-likelihood ratio (LLR) vector and a hard-bit estimate (HBE) vector. For a node $(s,t)$ of length~$\ell$, denote its LLR vector and HBE vector by $\boldsymbol{\alpha}^{(s,t)}=(\alpha^{(s,t)}_0,\ldots,\alpha^{(s,t)}_{\ell-1})$ and $\boldsymbol{\beta}^{(s,t)}=(\beta^{(s,t)}_0,\ldots,\beta^{(s,t)}_{\ell-1})$, respectively.

After initializing the LLRs at the root using the channel observations, the SC decoder proceeds according to the following three update rules.

\textbf{LLR update for the left child:}
When the LLRs of a parent node are available, the LLRs of its left child are computed. If $\ell$ is even,
\begin{align}
\alpha^{(s+1,2t)}_i
=
f\big(\alpha^{(s,t)}_i,\alpha^{(s,t)}_{\lceil\ell/2\rceil+i}\big),
\quad
0\le i\le \lceil\ell/2\rceil-1,
\end{align}
If $\ell$ is odd,
\begin{align}
\alpha^{(s+1,2t)}_i=
\begin{cases}
f(\alpha^{(s,t)}_i,\alpha^{(s,t)}_{\lceil\ell/2\rceil+i}),
&0\le i\le\lceil\ell/2\rceil-2\\
\alpha^{(s,t)}_i,
&i=\lceil\ell/2\rceil-1.
\end{cases}
\end{align}
Here, $f(a,b)=\ln\frac{1+e^{a+b}}{e^a+e^b}, a,b\in\mathbb{R}$.

\textbf{LLR update for the right child:}
Once the HBEs of the left child are available, the LLRs of the right child are computed as
\begin{align}
\alpha^{(s+1,2t+1)}_i
=
g\big(
\alpha^{(s,t)}_i,
\alpha^{(s,t)}_{\lceil\ell/2\rceil+i},
\beta^{(s+1,2t)}_i
\big),
\end{align}
for $0\le i\le\lfloor\ell/2\rfloor-1$, where $g(a,b,c)=b+(-1)^c a$.

\textbf{HBE update for the parent:} 
Given the HBEs of both children, the HBEs of the parent node are computed as
\begin{align}
\boldsymbol{\beta}^{(s,t)}
=
\big(
\boldsymbol{\beta}^{(s+1,2t)} \boxplus \boldsymbol{\beta}^{(s+1,2t+1)},
\boldsymbol{\beta}^{(s+1,2t+1)}
\big).
\end{align}

In contrast to the LLR propagation, which proceeds from the root downward, the HBEs are propagated from the leaves upward. Frozen leaves are assigned zero HBEs, while active leaves are decided via threshold detection: $\beta=1$ if $\alpha\le0$, and $\beta=0$ otherwise. After the HBEs of all nodes have been computed, the decoder extracts the estimated data bits from the active leaf nodes.
\begin{remark}
It is worth emphasizing that the BBT channel transformation, together with the resulting BBT polar codes, provides a natural and direct generalization of the original channel transformation and polar codes introduced in~\cite{Arikan2009Polarization}.
\end{remark}

\section{BBT Channel Polarization}\label{section3}
In this section, we show that the proposed BBT channel transformation induces channel polarization. Both theoretical analysis and numerical results are provided to illustrate this phenomenon.
\subsection{Theoretical Results}
Using the normal-graph representation of the BBT channel transformation, we establish the channel polarization theorem. To this end, we first recall several fundamental results from~\cite{Alsan2014PolarizationProof} and then adapt the combinatorial argument therein to the BBT channel transformation.
\begin{lemma}[\cite{Alsan2014PolarizationProof}~]
Let $W_0$ and $W_1$ be two independent BMS channels, and let $W_0 \ast W_1$ and $W_0 \circ W_1$ denote the corresponding ``upper'' and ``lower'' synthesized channels under the polar transformation. Then,
\begin{align}
I(W_0) + I(W_1) = I(W_0 \ast W_1) + I(W_0 \circ W_1).
\end{align}
In other words, the polar transformation preserves the symmetric capacity.
\end{lemma}

\begin{definition}
For constants $0 < a < b < 1$, a BMS channel $W$ is classified as good~(i.e., reliable), mediocre~(i.e., non-polarized), or bad~(i.e., unreliable) if $I(W)\in( b,1 ]$, $I(W)\in[a,b]$, or $I(W)\in[0,a)$, respectively. 	
\end{definition}

\begin{lemma}[\cite{Alsan2014PolarizationProof}~]
For any constants $0 < a < b < 1$, if $I\left(W_0\right),I\left(W_1\right)\in\left[a,b\right]$, then there exists a strictly positive function $\zeta(a,b)$ such that
\begin{align}
\left[I^2\left(W_0\ast W_1\right)+I^2\left(W_0\circ W_1\right)\right]-\left[I^2\left(W_0\right)+I^2\left(W_1\right)\right] \geq \zeta(a, b)>0.
\end{align}
In other words, when two mediocre channels are combined, the polar transformation strictly increases the dispersion of their symmetric capacities.
\end{lemma}

\begin{definition}
Consider a block of $\ell$ independent channels. The block is called extremal if no two mediocre channels are combined while the block undergoes the BBT polar channel transformations; otherwise, it is non-extremal.
\end{definition}

\begin{lemma}[\cite{Alsan2014PolarizationProof}~]
Let $\ell = 2^\delta~(\delta\geq 0)$. For any extremal block of $\ell$ independent channels, the number of mediocre channels in the block is at most $\binom{\delta}{\lfloor \delta/2 \rfloor}$.
\end{lemma}

We now generalize Lemma~3 to blocks of arbitrary length.

\begin{lemma}
Let $\ell$ be an arbitrary positive integer and let $\delta = \lceil \log_2 \ell \rceil$. For any extremal block of $\ell$ independent channels, the number of mediocre channels in the block is at most $\binom{\delta}{\lfloor \delta/2 \rfloor}$.
\end{lemma}
\begin{proof}
See Appendix.
\end{proof}

\begin{theorem}[BBT Channel Polarization]\label{BBTChannelPolarizationTheorem}
Let $W$ be a BMS channel. Consider $N$ independent copies of $W$ that undergo the BBT polar transformation. In the factor graph representation of the BBT transformation of length $N$, the $N$ channels $\{\widetilde W_{0,j}:0\le j\le N-1\}$ at layer $0$ produce $N$ synthetic channels $\{\widetilde W^{(N)}_{n,j}:0\le j\le N-1\}$ at layer $n$, where $n\triangleq \lceil \log_2 N\rceil$.

Then, for any constants $0<a<b<1$, the following limits hold:
\begin{align}
\lim_{N\to\infty}\frac{1}{N}
\#\bigl\{0\le j\le N-1:\, I(\widetilde W_{n,j}^{(N)})\in[0,a)\bigr\}
&=1-I(W), \\
\lim_{N\to\infty}\frac{1}{N}
\#\bigl\{0\le j\le N-1:\, I(\widetilde W_{n,j}^{(N)})\in[a,b]\bigr\}
&=0, \\
\lim_{N\to\infty}\frac{1}{N}
\#\bigl\{0\le j\le N-1:\, I(\widetilde W_{n,j}^{(N)})\in(b,1]\bigr\}
&=I(W),
\end{align}
where $I(\cdot)$ denotes the symmetric capacity of a channel and $\#$ denotes the cardinality of a set.
\end{theorem}

\begin{proof}
For each layer $m$, define
\begin{align}
\Omega_m(a,b)=\frac{1}{N}\#\left\{j\in\{0,1,\ldots,N-1\}:I(\widetilde W_{m,j})\in[a,b]\right\},
\end{align}
which represents the fraction of mediocre channels at layer $m$. Similarly, define
\begin{align}
\Psi_m(a)=\frac{1}{N}\#\left\{j\in\{0,1,\ldots,N-1\}:I(\widetilde W_{m,j})<a\right\},\\
\Phi_m(b)=\frac{1}{N}\#\left\{j\in\{0,1,\ldots,N-1\}:I(\widetilde W_{m,j})>b\right\}.
\end{align}

Since $n=\lceil\log_2 N\rceil$, letting $N\to\infty$ is equivalent to letting $n\to\infty$. For any fixed $\delta>0$, we consider layers $m=n-\delta$. As $n\to\infty$, we also have $m\to\infty$. Our goal is to show that, as $m\to\infty$, $\Omega_m(a,b)$, $\Psi_m(a)$, and $\Phi_m(b)$ converge to $0$, $1-I(W)$ and $I(W)$, respectively. To this end, define
\begin{align}
\mu_m=\frac{1}{N}\sum_{j=0}^{N-1} I(\widetilde W_{m,j}),\qquad
\nu_m=\frac{1}{N}\sum_{j=0}^{N-1} I^2(\widetilde W_{m,j}).
\end{align}

The proof proceeds in three steps.

\textbf{1)~Conservation of symmetric capacity:}  
By Lemma~1, each polar transformation preserves the sum of symmetric capacities. Summing over all polar transformations between layers $m$ and $m+1$ yields $\mu_{m+1}=\mu_m$. Hence
\begin{align}
\mu_m=\mu_0=I(W),\qquad \forall m\ge0.
\end{align}

\textbf{2)~Monotonicity of the second moment:}  
For any pair of mediocre channels $(W_0,W_1)$ with $I(W_0),I(W_1)\in[a,b]$ that are combined by a polar transformation, Lemma~2 implies that the sum of their squared symmetric capacities increases by at least $\zeta(a,b)>0$. Summing over all polar transformations between layers $m$ and $m+1$ yields $\nu_{m+1}\ge\nu_m$. Since $0\le\nu_m\le1$, the sequence $\{\nu_m\}$ is bounded and monotone, and hence convergent.

\textbf{3)~Vanishing fraction of mediocre channels:}  
It remains to show that $\lim_{m\to\infty}\Omega_m(a,b)=0$.

Suppose the BBT channel transformation for $N$ original BMS channels proceeds from layer $0$ to layer $m$ on a factor graph with $n$ layers, where $m<n$. According to the factor graph, the channels at layer $m$ can be partitioned into $2^m$ disjoint blocks. Let $\delta=n-m$. Each block contains at most $2^\delta$ channels and will subsequently undergo a BBT polar transform with $\delta$ stages.

According to Definition~3, these blocks can be classified as extremal or non-extremal. Let $\gamma_m$ denote the fraction of non-extremal blocks, and define $\epsilon_\delta=\binom{\delta}{\lfloor\delta/2\rfloor}/2^\delta$. By Lemma~4, the number of mediocre channels at layer $m$ satisfies
\begin{align}
N\Omega_m(a,b)
\le 2^m(1-\gamma_m)\binom{\delta}{\lfloor\delta/2\rfloor}
+2^m\gamma_m2^\delta
=
2^{m+\delta}\left(\left(1-\gamma_m\right) \epsilon_\delta+\gamma_m\right).
\end{align}
Since $1/2\le N/2^{m+\delta}\le1$, it follows that
\begin{align}
\Omega_m(a,b)\le2\left((1-\gamma_m)\epsilon_\delta+\gamma_m\right).
\end{align}

We now prove that $\lim_{m\to\infty}\Omega_m(a,b)=0$ by contradiction. Suppose $\Omega_m(a,b)>2\epsilon_\delta$. Then
\begin{align}
\gamma_m\ge
\frac{\Omega_m(a,b)/2-\epsilon_\delta}{1-\epsilon_\delta}>0,
\end{align}
so a strictly positive fraction of blocks is non-extremal.

During the BBT transformation from layer $m$ to layer $m+\delta$, each non-extremal block must contain at least one polar transformation combining two mediocre channels. By Lemma~2, this operation increases the squared symmetric capacity by at least $\zeta(a,b)$. Hence
\begin{align}
\nu_{m+\delta}-\nu_m
\ge
\frac{\gamma_m2^m\zeta(a,b)}{N}
\ge
\frac{\gamma_m\zeta(a,b)}{2^\delta}>0.
\end{align}
However, since $\{\nu_m\}$ is convergent, the difference $\nu_{m+\delta}-\nu_m$ must vanish as $m\to\infty$, yielding a contradiction. Therefore $\Omega_m(a,b)\le2\epsilon_\delta$. Since $\lim_{\delta\to\infty}\epsilon_\delta=0$, we conclude that
\begin{align}
\lim_{m\to\infty}\Omega_m(a,b)=0.
\end{align}

Finally, since for all $m$, we have
\begin{align}
I(W)&=\mu_m,\\
\mu_m&\leq a \Psi_m(a)+b \Omega_m(a, b)+\Phi_m(b)=a+(b-a) \Omega_m(a, b)+(1-a) \Phi_m(b),\\
1-\mu_m&\leq 1-a \Omega_m(a, b)-b\Phi_m(b)=\Psi_m(a)+(1- a) \Omega_m(a, b)+(1-b) \Phi_m(b).
\end{align}
By letting $a\to0$ in the second inequality, we obtain
\begin{align}
\liminf_{m\to\infty} \Phi_m(b)\geq I(W).
\end{align}
Similarly, by letting $b\to1$ in the third inequality, we obtain
\begin{align}
\liminf_{m\to\infty} \Psi_m(a)\geq 1-I(W).
\end{align}
Since $\Psi_m(a)+\Phi_m(b)\leq 1$, it follows that
\begin{align}
\lim_{m\to\infty} \Psi_m(a)=1-I(W), \qquad
\lim_{m\to\infty} \Phi_m(b)=I(W).
\end{align}
This completes the proof.
\end{proof}
\begin{corollary}
For any BMS channel $W$ and any coding rate $R<I(W)$, there exists a sequence of BBT polar codes with increasing block lengths whose information sets are selected from the most reliable synthetic channels induced by the BBT channel transformation, such that the block error probability under SC decoding tends to zero as $N\to\infty$.
\end{corollary}
\begin{proof}
From Theorem~1, the fraction of reliable synthetic channels approaches $I(W)$ as $N\to\infty$. Hence, for any $R<I(W)$, one can select an information set consisting only of sufficiently reliable synthetic channels. Under SC decoding, the block error probability is upper bounded by the sum of the error probabilities of the selected synthetic channels, which tends to zero as $N\to\infty$.
\end{proof}
\begin{remark}
The proof of Theorem~1 follows the elementary proof strategy of Alsan and Telatar~\cite{Arikan2009Polarization} in that it avoids martingale arguments and relies solely on elementary methods. However, the present result is not a direct restatement of~\cite{Alsan2014PolarizationProof}. The key novelty lies in extending this proof framework from the classical polar transformation with block length $2^n$ to the BBT channel transformation with arbitrary block length. 

Note that the BBT channel transformation involves irregular block sizes and odd-length splittings, under which the extremal block argument used in~\cite{Alsan2014PolarizationProof} no longer applies directly. This difficulty is resolved in Lemma~4 by introducing a new subset-labeling argument. Specifically, it is shown that for any extremal BBT block of length $\ell$, the number of mediocre channels remains bounded by a Sperner-type quantity $\binom{\delta}{\lfloor \delta/2 \rfloor}$ with $\delta=\lceil \log_2 \ell \rceil$. This combinatorial generalization is the key step that enables the polarization theorem for BBT transformations and consequently establishes the capacity-achieving property of BBT polar codes for arbitrary block lengths.
\end{remark}
\subsection{Numerical Results}
\begin{figure}[t]
\centering
\includegraphics[width=0.68\textwidth]{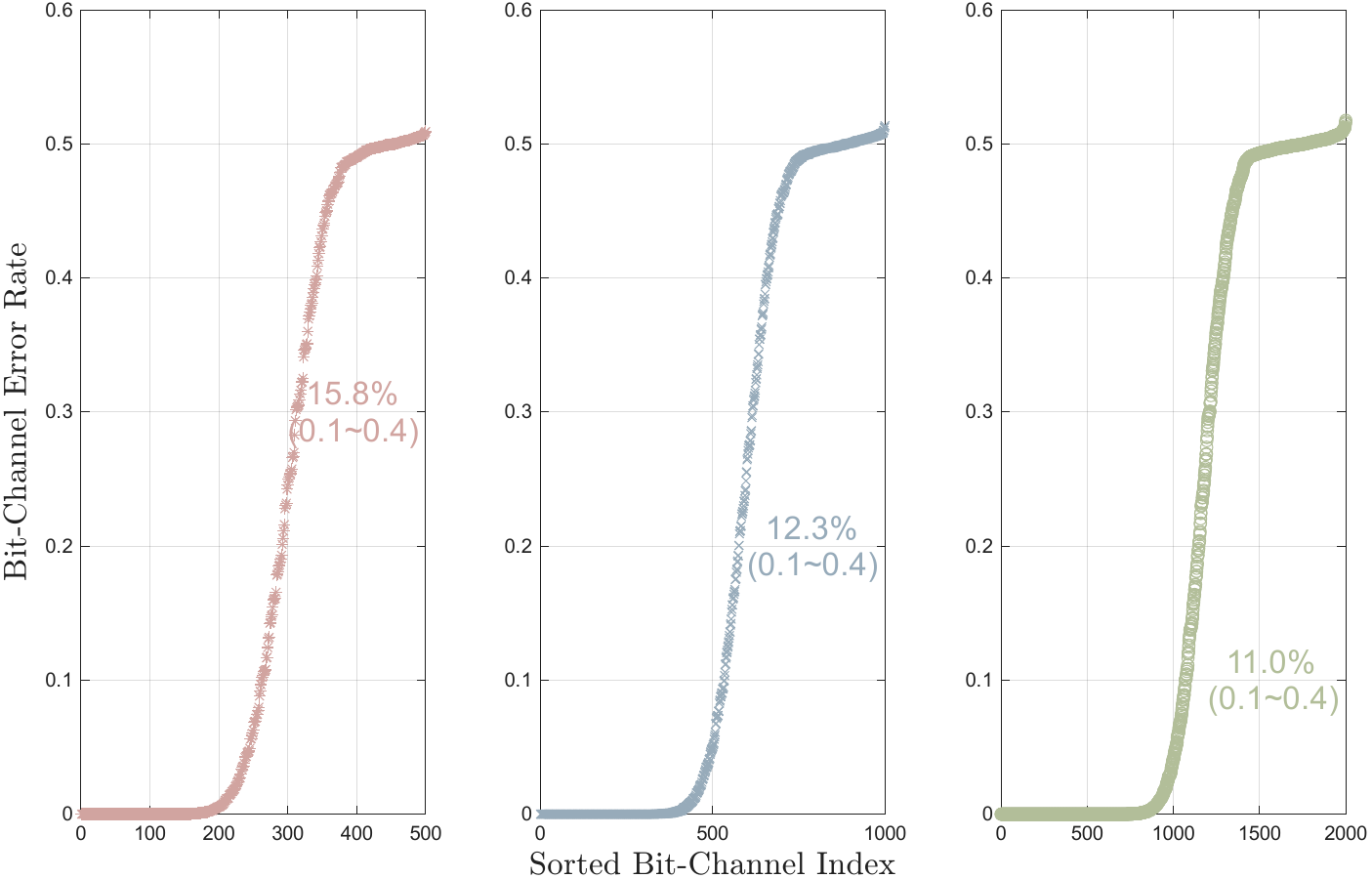}
\caption{Sorted bit-channel error rates for BBT polar codes with $N = 500$, $1000$, and $2000$ at SNR $= 1$ dB. The percentage values indicate the fraction of bit-channels with error rates in the interval $[0.1, 0.4]$.}
\label{PolarizationEffectOverIBBT}
\end{figure}
To validate the polarization phenomenon induced by the BBT channel transformation, we perform Monte Carlo simulations over binary phase-shift keying-additive white Gaussian noise~(BPSK-AWGN) channels at a signal-to-noise ratio (SNR) of $1$~dB. The error rate of each synthesized bit-channel is estimated based on $10^4$ independent trials.

Fig.~\ref{PolarizationEffectOverIBBT} presents the sorted bit-channel error rates for block lengths $N = 500$, $1000$, and $2000$, where the synthesized bit-channels are ordered in increasing order of their error rates. For all block lengths, the majority of bit-channels exhibit error rates either below $0.1$ or above $0.4$, indicating a clear polarization effect produced by the BBT transformation. Moreover, as the block length increases from $500$ to $2000$, the fraction of bit-channels with intermediate error rates in the interval $[0.1, 0.4]$ decreases from $15.8\%$ to $12.3\%$, and further to $11.0\%$. This trend is consistent with the channel polarization behavior predicted by Theorem~1.

\section{Weight Spectrum and Performance Bounds}\label{section4}
In this section, we investigate the ML performance of BBT polar codes. Our analysis proceeds in two steps. First, by introducing the random permutations, we develop an estimation method for the weight spectrum of BBT polar codes. Then, we introduce efficient upper and lower bounds on the FER of linear block codes under ML decoding. Together, these tools enable the evaluation of the ML performance of any given BBT polar code.

\subsection{WEF Estimation for BBT Polar Codes}
\begin{figure}[t]
\centering
\includegraphics[width=0.5\textwidth]{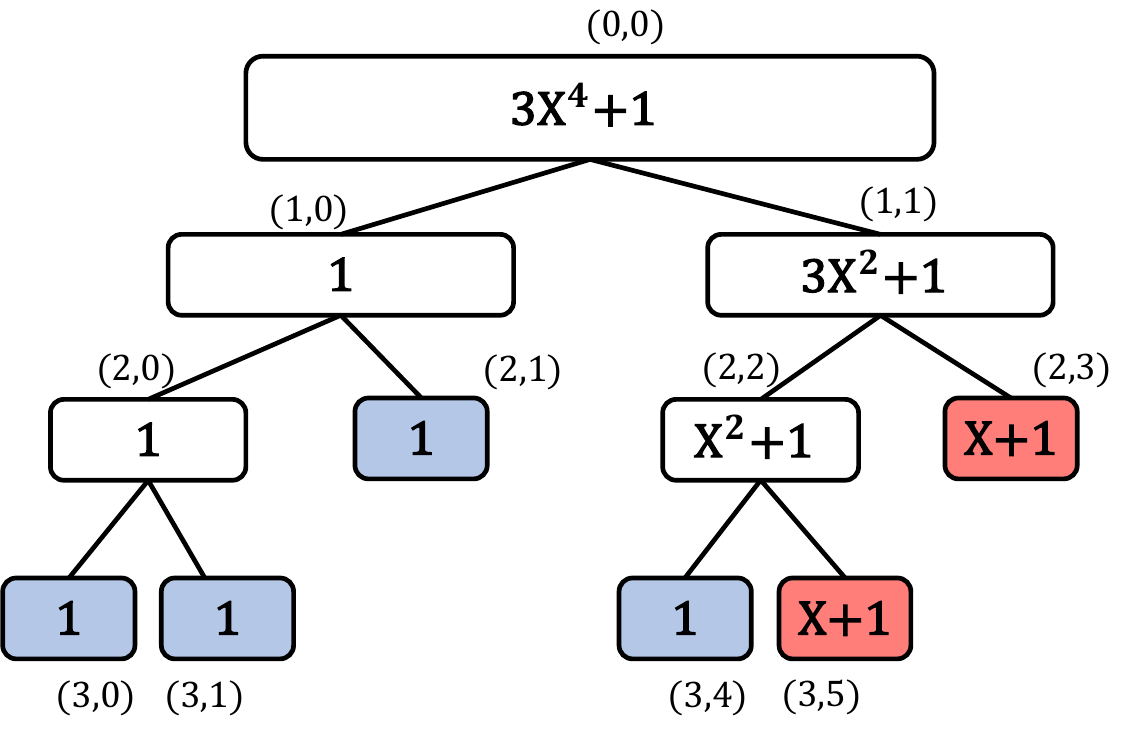}
\caption{Average WEFs of BBT tree nodes for $\mathcal{C}_{\mathrm{BBT}}[6,2,\{4,5\}]$.}
\label{BBTWEFForN6}
\end{figure}
We begin by presenting a theorem that generalizes the result in~\cite{Chiu2020IPolar,Ma2025PlotkinWeightDistribution} and forms the basis for estimating the WEFs of BBT polar codes.
\begin{theorem}\label{WEFOfCompositeCode}
Let $\mathcal{C}^{\ast}$ and $\mathcal{C}^{\circ}$ be two binary linear block codes of lengths $\ell^{\ast}$ and $\ell^{\circ}$, where $\ell^{\ast} \ge \ell^{\circ}$. Consider the composite binary linear block code ensemble of length $\mathcal{C}$ defined as
\begin{align}\label{CompositeCodeEnsemble}
\mathcal{C}=
\left\{\left(\boldsymbol{v}^{\ast} \boxplus (\boldsymbol{v}^{\circ},\boldsymbol{0}^{\ell^{\ast}-\ell^{\circ}})\Pi,\,\boldsymbol{v}^{\circ}\right)\,\middle|\,\boldsymbol{v}^{\ast}\in\mathcal{C}^{\ast},\,\boldsymbol{v}^{\circ}\in\mathcal{C}^{\circ}\right\},
\end{align}
where $\Pi$ is a uniformly distributed random permutation matrix. If the WEFs of $\mathcal{C}^{\ast}$ and $\mathcal{C}^{\circ}$ are $A^{\ast}(X)=\sum_{w=0}^{\ell^{\ast}} A^{\ast}_{w} X^w$ and $A^{\circ}(X)=\sum_{w=0}^{\ell^{\circ}} A^{\circ}_{w} X^w$, respectively, then the average WEF of $\mathcal{C}$ is
\begin{align}
A(X)=\sum_{w=0}^{\ell} A_w X^w,
\end{align}
where
\begin{align}\label{Aw}
A_w
&=
\sum_{i=\max\{0,w-\ell^{\ast}\}}^{\min\{w,\ell^{\circ}\}}
\sum_{j=\max\{0,w-\ell^{\ast}\}}^{\min\{i,w-i\}}
\frac{\binom{\ell^{\ast}}{i}\binom{w-i}{j}\binom{\ell^{\ast}-i}{i-j}}
{\binom{\ell}{i}\,\binom{\ell^{\ast}}{w-2j}}
A^{\circ}_{i}\, A^{\ast}_{w-2j}.
\end{align}
\end{theorem}
\begin{proof}
Let $w\le \ell$ and define the Hamming sphere
\begin{align}
\mathcal{S}^{\ell}_w=\{\boldsymbol{v}\in\mathbb{F}_2^\ell: wt(\boldsymbol{v})=w\}, 
\end{align}
where $wt(\boldsymbol{v})$ denotes the Hamming weight of $\boldsymbol{v}$. Let $\boldsymbol{V}$ be uniformly distributed over $\mathcal{S}^{\ell}_w$, and then we have
\begin{align}\label{ProbCalculation1}
\Pr\{\boldsymbol{V}\in\mathcal{C}\} = \frac{A_w}{\binom{\ell}{w}}.
\end{align}	
Now, we write $\boldsymbol{V}=(\boldsymbol{V}^{\ast},\boldsymbol{V}^{\circ})$ and define $\widetilde{\boldsymbol{V}^{\ast}}= \boldsymbol{V}^{\ast} \boxplus (\boldsymbol{V}^{\circ},\boldsymbol{0}^{\ell^{\ast}-\ell^{\circ}})\Pi$, where $\boldsymbol{V}^{\ast}\in\mathbb{F}_2^{\ell^{\ast}}$ and $\boldsymbol{V}^{\circ}\in\mathbb{F}_2^{\ell^{\circ}}$. Since $\Pi$ is uniformly distributed, we have
\begin{align}\label{ProbCalculation2}
\Pr\{\boldsymbol{V}\in\mathcal{C}\}
=
\sum_{i=\max \{0,w-\ell^{\ast}\}}^{\min \{w,\ell^{\circ}\}}
\frac{\binom{\ell^{\circ}}{i}\binom{\ell^{\ast}}{w-i}}{\binom{\ell}{w}}\cdot \frac{A^{\circ}_{i}}{\binom{\ell^{\circ}}{i}} 
\quad \times
\sum_{j=\max \{0,w-\ell^{\ast}\}}^{\min \{i,w-i\}}
\frac{\binom{w-i}{j}\binom{\ell^{\ast}-w+i}{i-j}}{\binom{\ell^{\ast}}{i}}
\cdot \frac{A^{\ast}_{w-2j}}{\binom{\ell^{\ast}}{w-2j}} .
\end{align}
By equating \eqref{ProbCalculation1} and \eqref{ProbCalculation2}, we can obtain \eqref{Aw}.
\end{proof}

For a given BBT polar code $\mathcal{C}_{\mathrm{BBT}}[N,K,\mathcal{A}]$, determining its exact WEF is generally computationally intractable. Nevertheless, during the encoding process the code vector at each parent node is obtained from those of its children according to $\boldsymbol{v}^{(s,t)}=
\left(\boldsymbol{v}^{(s+1,2t)} \boxplus \boldsymbol{v}^{(s+1,2t+1)},\boldsymbol{v}^{(s+1,2t+1)}\right)$, which can match the composite code construction in Theorem~\ref{WEFOfCompositeCode}. This observation enables an efficient WEF estimation procedure.

Given the code length $N$ and the information set $\mathcal{A}$, we first construct the corresponding BBT-structured tree graph and initialize the WEFs of the leaf nodes: $A(X)=1+X$ for active leaves and $A(X)=1$ for frozen leaves. The WEFs of all internal nodes are then estimated recursively in a bottom-up manner using Theorem~\ref{WEFOfCompositeCode}. This procedure yields an estimate of the WEF of the BBT polar code at the root node. As an illustrative example, the estimated WEF of $\mathcal{C}_{\mathrm{BBT}}[6,2,\{4,5\}]$ is shown in Fig.~\ref{BBTWEFForN6}.
\subsection{MHW Spectrum Analysis}
\label{subsec:MinWeightSpectrum}

In many practical scenarios, the ML decoding performance of a linear block code is primarily governed by its non-zero MHW and the multiplicity of MHW codewords. In this subsection, we develop an efficient recursive method for computing the MHW spectrum of BBT polar codes. 
\subsubsection{State Representation}
For each node in the BBT encoding tree, the associated constituent code can be classified into two types. One is zero-only subcodes, which contain only the all-zero codeword, and the other is non-zero subcodes, which contain at least one non-zero codeword.

Accordingly, each node is associated with a binary indicator $\mathrm{IsNonZero}\in\{0,1\}$, which specifies whether the corresponding subcode contains any non-zero codeword. If $\mathrm{IsNonZero}=1$, the node is further characterized by a pair $(w_{\min},A_{w_{\min}})$, where $w_{\min}$ denotes the non-zero MHW and $A_{w_{\min}}$ denotes the number of codewords attaining this weight. At the leaf level, nodes carrying data bits are initialized as non-zero subcodes with $\mathrm{IsNonZero}=1$ and $(w_{\min},A_{w_{\min}})=(1,1)$, while frozen nodes correspond to zero-only subcodes with $\mathrm{IsNonZero}=0$.

\subsubsection{MHW Recursion}
Consider an internal node $(s,t)$ whose left and right children are $(s+1,2t)$ and $(s+1,2t+1)$, respectively. According to the BBT encoding rule
\begin{align}
\boldsymbol{v}^{(s,t)}
=
\big(
\boldsymbol{v}^{(s+1,2t)} \boxplus \boldsymbol{v}^{(s+1,2t+1)},
\boldsymbol{v}^{(s+1,2t+1)}
\big),
\end{align}
the minimum weight of the parent node depends on the spectra of its two children. If exactly one child contains non-zero codewords, the MHW of the parent is inherited from that child:
\begin{align}
w_{\min}^{(s,t)} =
\begin{cases}
w_{\min}^{(s+1,2t)}, & \text{if the right child is zero-only},\\
2w_{\min}^{(s+1,2t+1)}, & \text{if the left child is zero-only}.
\end{cases}
\end{align}
If both children contain non-zero codewords, Theorem~2 shows that the MHW of the parent satisfies
\begin{align}
w_{\min}^{(s,t)}
=
\max
\left\{
w_{\min}^{(s+1,2t)},
\,
2w_{\min}^{(s+1,2t+1)}-w_{\min}^{(s+1,2t)}
\right\}.
\label{eq:MHW_parent}
\end{align}
Hence, the MHW of the parent node depends solely on the MHWs of its two children.

\subsubsection{Multiplicity Recursion}
The multiplicity of MHW codewords follows directly from the structure of the BBT combination.

If exactly one child contains non-zero codewords, the multiplicity is inherited from that child:
\begin{align}
A_{w_{\min}^{(s,t)}} =
\begin{cases}
A_{w_{\min}^{(s+1,2t)}}, & \text{if the right child is zero-only},\\
A_{w_{\min}^{(s+1,2t+1)}}, & \text{if the left child is zero-only}.
\end{cases}
\end{align}
If both children contain non-zero codewords, two cases arise:
\begin{itemize}
\item If $w_{\min}^{(s+1,2t)} \neq w_{\min}^{(s+1,2t+1)}$, the MHW of the parent  arises from a unique cancellation pattern, yielding
\begin{align}
A_{w_{\min}^{(s,t)}} =
A_{w_{\min}^{(s+1,2t)}}
A_{w_{\min}^{(s+1,2t+1)}}.
\end{align}
\item If $w_{\min}^{(s+1,2t)} = w_{\min}^{(s+1,2t+1)}$, two symmetric constructions achieve the MHW, resulting in
\begin{align}
A_{w_{\min}^{(s,t)}} =
2
A_{w_{\min}^{(s+1,2t)}}
A_{w_{\min}^{(s+1,2t+1)}}.
\end{align}
\end{itemize}
\subsubsection{Recursive Spectrum Computation}
Combining the above rules, the MHW spectrum can be computed recursively in a bottom-up traversal of the BBT encoding tree. Starting from the leaf initialization, the recursion propagates the MHW and its multiplicity to the root node. The resulting procedure is summarized in Algorithm~\ref{alg:min_weight_bbt}.
\begin{algorithm}[!h]
\caption{Recursive Computation of the MHW Spectrum for BBT Polar Codes}
\label{alg:min_weight_bbt}

\KwIn{BBT structure of length $N$; information set $\mathcal{A}$}

\KwOut{MHW $w_{\min}$ and multiplicity $A_{w_{\min}}$}

Construct the BBT encoding tree of length $N$\;

\ForEach{leaf node $i=0,1,\ldots,N-1$}{
\uIf{$i\in\mathcal{A}$}{
$\mathrm{IsNonZero}(i)\leftarrow1, (w_{\mathrm{leaf},i},A_{w_{\mathrm{leaf},i}})\leftarrow(1,1)$;
}
\Else{
$\mathrm{IsNonZero}(i)\leftarrow0$;
}
}

\ForEach{parent node $p$ in bottom-up order}{		
Let $l$ and $r$ denote the left and right children of the node $p$\;
\uIf{$\mathrm{IsNonZero}(l)=0$ \textbf{and} $\mathrm{IsNonZero}(r)=0$}{
$\mathrm{IsNonZero}(p)\leftarrow0$;
}		
\uElseIf{$\mathrm{IsNonZero}(l)=1$ \textbf{and} $\mathrm{IsNonZero}(r)=0$}{
$\mathrm{IsNonZero}(p)\leftarrow1, (w_p,A_{w_p})\leftarrow(d_l,A_{d_l})$;
}

\uElseIf{$\mathrm{IsNonZero}(l)=0$ \textbf{and} $\mathrm{IsNonZero}(r)=1$}{
$\mathrm{IsNonZero}(p)\leftarrow1,(w_p,A_{w_p})\leftarrow(2w_r,A_{w_r})$;
}
\Else{			
$\mathrm{IsNonZero}(p)\leftarrow1, w_p \leftarrow \max(w_l,2w_r-w_l)$\;		
\uIf{$w_l=w_r$}{
$A_{w_p}\leftarrow 2A_{w_l}A_{w_r}$\;
}
\Else{
$A_{w_p}\leftarrow A_{w_l}A_{w_r}$\;
}
}
}

\Return{$(w_{\min},A_{w_{\min}})=(w_{\mathrm{root}},A_{w_{\mathrm{root}}})$}
\end{algorithm}

\subsection{ML Upper Bound and Lower Bound}
Consider the ML decoding of a binary linear block code with WEF $A(X)=\sum_{w} A_w X^w$. An analytical upper bound on the FER was derived in~\cite{Ma2013MLBound}, which is summarized as follows. Define
\begin{align}
Q(x)=\frac{1}{\sqrt{2\pi}}\int_x^\infty e^{-z^2/2}\,dz,\quad p_b=Q(1/\sigma),
\end{align}
and
\begin{align}
e_0(p,N_t,N_\ell,N_u)=\sum_{m=N_\ell}^{N_u}\binom{N_t}{m}p^m(1-p)^{N_t-m}.
\end{align}
Furthermore, define
\begin{align}
e_1(A_w)
&=
A_w\, Q(\sqrt{w}/\sigma)\, e_0(p_b,N-w,0,\lfloor w^\ast-w/2\rfloor),\\
e_2(A_w)
&=
(A_w-1)
\left(Q(\sqrt{w}/\sigma)-\tfrac12 Q^2(\sqrt{w}/\sigma)\right)
e_0(p_b,N-2w,0,\lfloor w^\ast-w/2\rfloor)
+
Q(\sqrt{w}/\sigma).
\end{align}
Then, the FER upper bound is given by
\begin{align}\label{MLUB}
\mathrm{FER_{UB}}=\min_{0\le w^\ast\le N}\left\{\sum_{w\le 2w^\ast} \min\{e_1(A_w),e_2(A_w)\}+e_0(p_b,N,w^\ast+1,N)\right\}.
\end{align}

On the other hand, based on the Kuai-Alajaji-Takahara~(KAT) bound~\cite{Kuai2000UnionProbabilityLowerBound}, an efficient algorithmic lower bound on the FER was proposed in~\cite{Behnamfar2007LinearCodeErrorBound}.

\begin{theorem}[KAT Bound]
For a linear block code $\mathcal{C}$ of length $N$ with non-zero MHW $w_{\min}$, let $\mathcal{I}\subseteq\mathcal{C}$ be a subset with WEF $A(\mathcal{I};X)=\sum_{w=1}^N A_w(\mathcal{I})X^{w}$, where $A_w(\mathcal{I})$ denotes the number of codewords in $\mathcal{I}$ with Hamming weight $w$. The KAT lower bound is determined by the WEF and given by
\begin{align}
\mathrm{KAT}\left(A(\mathcal{I};X)\right)=
\sum_{w=1}^N
A_w(\mathcal{I})
Q^2(\sqrt{w}/\sigma)
\left(
\frac{\tilde{\theta}_w}{(2-\tilde{\theta}_w)Q(\sqrt{w}/\sigma)+\tilde{\beta}_w}
+
\frac{1-\tilde{\theta}_w}{(1-\tilde{\theta}_w)Q(\sqrt{w}/\sigma)+\tilde{\beta}_w}
\right),
\end{align}
where
\begin{align}
\tilde{\theta}_w
&=\frac{\tilde{\beta}_w}{\tilde{\alpha}_w}
-\left\lfloor\frac{\tilde{\beta}_w}{\tilde{\alpha}_w}\right\rfloor,\\
\tilde{\alpha}_w
&=Q(\sqrt{w}/\sigma),\\
\tilde{\beta}_w
&=(A_w(\mathcal{I})-1)
\Psi\!\left(1-\frac{w_{\min}}{2w},\sqrt{w}/\sigma,\sqrt{w}/\sigma\right)
+\sum_{t\ne 0,w}
A_t(\mathcal{I})\,\Psi(\kappa_{wt},\sqrt{w}/\sigma,\sqrt{t}/\sigma),
\end{align}
with
\begin{align}
\kappa_{wt}
=\min\left\{\frac{w}{t},\frac{t}{w},
\frac{w+t-w_{\min}}{2\sqrt{wt}}\right\},
\end{align}
and
\begin{align}
\Psi(\rho,x,y)
=
\frac{1}{2\pi\sqrt{1-\rho^2}}
\int_x^\infty \int_y^\infty
\exp\!\left(
-\frac{z^2 - 2\rho z t + t^2}{2(1-\rho^2)}
\right)
dz\, dt.
\end{align}
\end{theorem}
\begin{algorithm}[!t]
\caption{Iterative FER Lower Bound Computation}
\label{MLLB}
\KwIn{WEF of a binary linear block code $A(X)=\sum_{w} A_w X^w$, noise variance $\sigma^2$}
\KwOut{$\mathrm{FER_{LB}}$}
Initialize $A(\mathcal{I};X)\gets A_{w_{\min}}X^{w_{\min}}$\;
Compute $\mathrm{KAT}_{\text{best}} \gets \mathrm{KAT}(A(\mathcal{I};X))$\;
Set $w \gets w_{\min} + 1$\;
\While{$w \leq N$}{
$A(\mathcal{I};X) \gets A(\mathcal{I};X)+A_w X^w$\;
Compute $\mathrm{KAT}_{\text{new}} \gets \mathrm{KAT}(A(\mathcal{I};X))$\;
\If{$\mathrm{KAT}_{\text{new}} < \mathrm{KAT}_{\text{best}}$}{
\textbf{break}\;
}
$\mathrm{KAT}_{\text{best}} \gets \mathrm{KAT}_{\text{new}}$\;
$w \gets w + 1$\;
}
$\mathrm{FER_{LB}}\gets \mathrm{KAT}_{\text{best}}$\;
\Return{$\mathrm{FER_{LB}}$}\;
\end{algorithm}

The tightest KAT lower bound is given by $\max_{\mathcal{I}\subseteq\mathcal{C}}\mathrm{KAT}(\mathcal{I})$, which is generally intractable to compute due to the exponential number of possible subsets. However, a practical approximation can be obtained through an iterative algorithm that progressively enlarges the codeword subset $\mathcal{I}$ in order of increasing Hamming weight. The procedure is formalized in Algorithm~\ref{MLLB} and operates as follows: starting with the set of MHW codewords, we iteratively add all codewords of the next smallest weight and evaluate the KAT bound. The algorithm terminates when adding more codewords causes the bound to decrease, indicating that the maximum has been reached. This greedy approach typically yields a tight lower bound with reasonable computational complexity.

\subsection{Numerical Results}
\begin{figure*}[t]
\centering
\includegraphics[width=\textwidth]{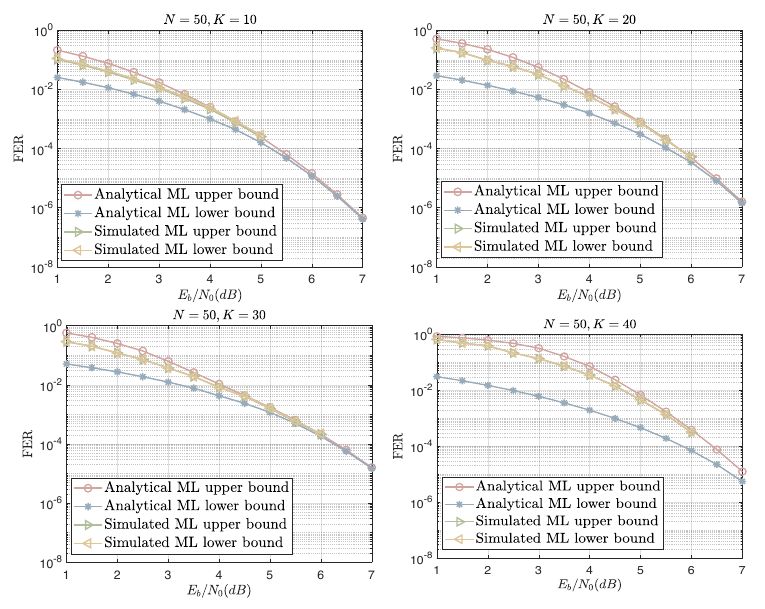}
\vskip-1.5em
\caption{FER performance of BBT polar codes with $N=50$ and $K=10, 20, 30,$ and $40$ under ML decoding. }
\label{fig:ML_performance}
\end{figure*}

In this subsection, the ML performance of BBT polar codes\footnote{In this work, all BBT polar codes are constructed using the universal polarization weight~(PW) method proposed in~\cite{Yao2024BalancedTree}.} is investigated through Monte Carlo simulations and analytical bounds. We consider BBT polar codes with block length $N=50$ and dimensions $K=10, 20, 30,$ and $40$. The analytical ML UB and LB, together with the simulated ML UB and LB, are shown in Fig.~\ref{fig:ML_performance}. The analytical ML UB and LB are computed according to \eqref{MLUB} and Algorithm~\ref{MLLB}, respectively. The simulated ML UB is obtained using ordered-statistics decoding~(OSD) in~\cite{Fossorier1995OSD}, while the simulated ML LB is obtained using the Chase-type soft-decision decoding algorithm in~\cite{Tang2022ChaseTypeRSDecoding}.

It can be observed that the simulated ML UB and the simulated ML LB are closely aligned, thereby providing an accurate estimate of the true ML performance. Moreover, in the high SNR region, the analytical ML upper and lower bounds are tightly matched and closely follow the true ML performance. These results confirm the effectiveness of the proposed ML bound computation formulas based on the average WEF, which enable reliable performance prediction of BBT polar codes.

\section{IBBT Polar Codes with BP Decoding}\label{section5}
In this section, we first construct the IBBT normal graph by modifying the BBT normal graph, and then derive the corresponding IBBT polar codes. Based on this graphical representation, we further develop a BP decoding algorithm that operates on a sub-graph of the IBBT normal graph.

\subsection{IBBT Normal Graph}
Recall that the normal graph representation of the BBT channel transformation can also be used to describe BBT polar codes. Motivated by this representation, we introduce a modified construction in which an interleaver is incorporated into the relationship between the parent and its children. Specifically, before combining the two child code vectors at an internal node, the right-child vector is permuted by an interleaver. When the lengths of the two child vectors are unequal, zero padding is applied prior to interleaving to ensure length consistency. This modification preserves the balanced-tree structure of the BBT construction while introducing controlled randomness into the recursive combination process.

Formally, for an internal node $(s,t)$ with length $\ell$, let the associated code vector be $\boldsymbol{v}^{(s,t)} \in \mathbb{F}_2^{\ell}$, and let its left and right children be $\boldsymbol{v}^{(s+1,2t)} \in \mathbb{F}_2^{\lceil \ell/2 \rceil}$ and $\boldsymbol{v}^{(s+1,2t+1)} \in \mathbb{F}_2^{\lfloor \ell/2 \rfloor}$, respectively. The node relation in the IBBT is defined as
\begin{equation}
\boldsymbol{v}^{(s,t)}
=
\big(
\boldsymbol{v}^{(s+1,2t)} \boxplus \boldsymbol{\tilde{v}}^{(s+1,2t+1)},\;
\boldsymbol{v}^{(s+1,2t+1)}
\big),
\end{equation}
where
\begin{equation}
\boldsymbol{\tilde{v}}^{(s+1,2t+1)}
\triangleq
\begin{cases}
\boldsymbol{v}^{(s+1,2t+1)} \Pi, 
& \text{if } \lceil \ell/2 \rceil = \lfloor \ell/2 \rfloor, \\[3pt]
(\boldsymbol{v}^{(s+1,2t+1)}, 0)\Pi, 
& \text{otherwise},
\end{cases}
\end{equation}
where $\Pi$ denotes a permutation matrix corresponding to the interleaver.

The resulting IBBT normal graph retains the same layered structure as the BBT normal graph, consisting of $n+1$ layers and $N$ variable nodes per layer, where $n=\lceil \log_2 N\rceil$. Each local connection between adjacent layers is represented by a pair of functional nodes, while the interleaver modifies the correspondence between variable nodes across layers. This modification may improve the connectivity of the normal graph and mitigate short cycles, which is beneficial for iterative BP decoding.

\subsection{IBBT Polar Codes}
For a given IBBT normal graph of length $N$, the mapping from the variable nodes at the bottom layer to those at the top layer can be represented by an invertible transformation matrix $\mathbf{G}_{\mathrm{IBBT}} \in \mathbb{F}_2^{N\times N}$. An IBBT polar code with code length $N$, dimension $K$, and information set $\mathcal{A} \subseteq \{0,1,\ldots,N-1\}$ is defined as
\begin{equation}
\mathcal{C}_{\mathrm{IBBT}}[N,K,\mathcal{A}]
=
\left\{
\mathbf{x}\mathbf{G}_{\mathrm{IBBT}}
\;\middle|\;
\mathbf{x}_{\mathcal{A}}=\mathbf{u} \in \mathbb{F}_2^K,\;
\mathbf{x}_{\mathcal{A}^c}=\mathbf{0}
\right\},
\end{equation}
where $\mathbf{u}$ denotes the information vector and $\mathbf{x}$ denotes the vector of bits assigned to the variable nodes at the bottom layer of the IBBT normal graph. The encoded vector $\mathbf{x}\mathbf{G}_{\mathrm{IBBT}}$ corresponds to the bits at the top layer of the graph and forms the transmitted codeword.

\subsection{BP Decoding on a Sub-Normal Graph}
IBBT polar codes preserve essentially the same code definition as BBT polar codes. Consequently, both encoding and BP decoding can be performed directly on the IBBT factor graph following procedures similar to those used for BBT polar codes, with the only difference being the topology of the underlying graph.

Let $\tau$ denote the truncation parameter with $0 \le \tau \le n$, chosen such that all nodes at the $(n-\tau)$-th layer correspond to IBBT subcodes whose lengths do not exceed $\ell_{\max}$. The resulting sub-normal graph spans the layers from the $0$-th to the $(n-\tau)$-th layer, thereby significantly reducing the number of layers involved in each iteration. The decoding procedure is summarized below.

\textbf{Initialization:}  
Given the received vector $\boldsymbol{y}$ and the information set $\mathcal{A}$, the downward messages at the $0$-th layer $\{D_{0,j}^{0}\}_{j=0}^{N-1}$ and the upward messages at the $n$-th layer $\{U_{n,j}^{0}\}_{j=0}^{N-1}$ are initialized in the same manner as in conventional BP decoding and remain fixed for all subsequent iterations. Starting from the $n$-th layer, the upward messages are propagated once to the $(n-\tau)$-th layer according to the standard BP update rules, producing the initial messages $\{U_{n-\tau,j}^{0}\}_{j=0}^{N-1}$.

\textbf{Iterative message passing:}  
For iterations $\kappa \ge 0$, the upward messages $\{U_{i,j}^{\kappa+1}\}$ are first updated from the $(n-\tau-1)$-th layer up to the $0$-th layer according to the standard BP update rules. Subsequently, the downward messages $\{D_{i,j}^{\kappa+1}\}$ are updated from the $1$-th layer down to the $(n-\tau)$-th layer.

Different from the conventional BP decoding algorithm, at the $(n-\tau)$-th layer, for each root node $(n-\tau,t)$ corresponding to an IBBT subcode $\mathcal{C}^{(n-\tau,t)}$ of length $\ell \le \ell_{\max}$, the upward messages associated with its variable nodes $\{[n-\tau,j+p]\}_{p=0}^{\ell-1}$ are updated directly as\footnote{The subtraction of $D_{n-\tau,j+p}^{\kappa}$ ensures that the resulting message is extrinsic with respect to the incoming information, thereby preventing double counting.}
\begin{equation}
U_{n-\tau,j+p}^{\kappa+1}
=
L_{p}^{(0)} - L_{p}^{(1)} - D_{n-\tau,j+p}^{\kappa},
\end{equation}
where
\begin{align}
L_{p}^{(b)}
&=
\ln
\!\!\!\!\!\!\!\!\!\!\!\!
\sum_{\boldsymbol{c}\in\mathcal{C}^{(n-\tau,t)}:c_p=b}
\!\!\!\!\!\!\!\!
\exp\!\left(
\frac{1}{2}\sum_{q=0}^{\ell-1}(1-2c_q)D_{n-\tau,j+q}^{\kappa}
\right),
\end{align}
for $b\in\{0,1\}$ and $p=0,1,\ldots,\ell-1$.

\textbf{Termination:}  
After each iteration, the posterior log-likelihood ratio at node $[i,j]$ is computed as $D_{i,j}^{\kappa}+U_{i,j}^{\kappa}$, based on which the decoding decision is obtained as
\begin{align}
\widehat{B}_{i,j}^{\kappa}
=
\mathrm{sign}\!\left(D_{i,j}^{\kappa}+U_{i,j}^{\kappa}\right).
\end{align}
If all sum-node and equality-node constraints in the sub-normal graph are satisfied, decoding terminates early. Otherwise, the iterations continue until the predefined maximum number of iterations $I_{\max}$ is reached.

\begin{remark}
The proposed BP decoding algorithm does not modify the code definition or the decoding objective, but only alters the message-passing schedule and the local processing strategy. When $\tau=0$, the sub-normal graph coincides with the full normal graph, and the proposed decoding algorithm reduces to conventional BP decoding.
\end{remark}

\begin{figure*}[t]
\centering
\includegraphics[width=0.8\textwidth]{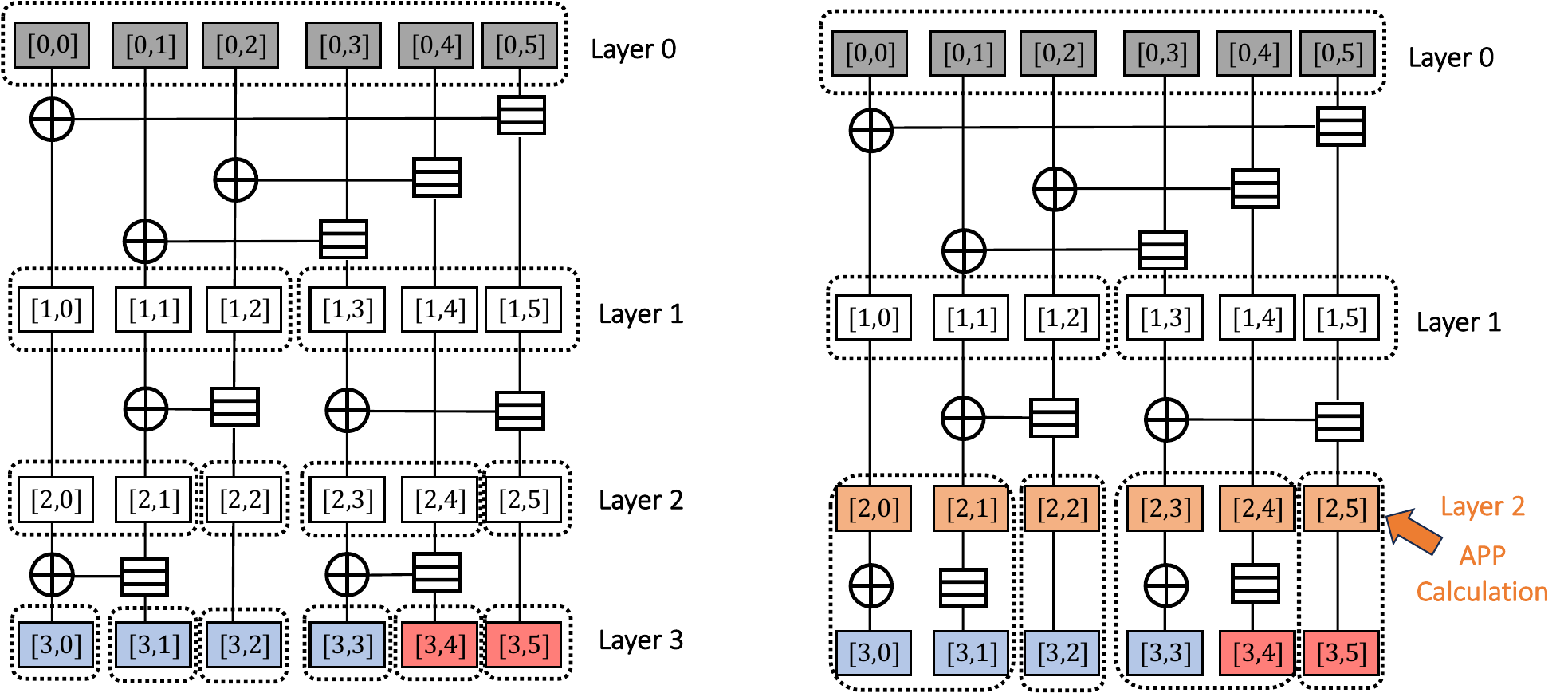}
\caption{Normal graphs of the IBBT polar code $\mathcal{C}_{\rm IBBT}[6,3,\{4,5\}]$.
Left: the original normal graph, where blue variable nodes correspond to frozen bits fixed to zero and the remaining nodes are active. 
Right: the sub-normal graph with $\ell_{\max}=2$, where iterative message passing is restricted to layers $0$ to $2$, and the APP calculation is performed at layer $2$.}
\label{IBBTSubNormalGraphs}
\end{figure*}

\begin{example}
Fig.~\ref{IBBTSubNormalGraphs} illustrates the code $\mathcal{C}_{\rm IBBT}[6,3,\{4,5\}]$. The left subfigure shows the original full normal graph, where the blue nodes are frozen (fixed to zero) and the remaining nodes are active. The right subfigure shows the corresponding sub-normal graph with $\ell_{\max}=2$, in which message passing is performed only across layers $0$ to $2$, while the APP calculation is carried out at layer $2$.
\end{example}

\subsection{Latency and Computational Complexity Analysis}
When BP decoding is implemented on the IBBT sub-normal graph, both the decoding latency and the computational complexity are jointly determined by the number of iterations, the number of message-passing steps per iteration, and the APP calculation modules. To characterize these aspects, we adopt the following implementation-oriented metrics.

\subsubsection{Latency Metrics}
As an iterative algorithm, the BP decoder terminates after a number of iterations that depends on the channel realization and the structure of the sub-normal graph. In addition, within each iteration, message passing is performed layer by layer along the graph.

Accordingly, the decoding latency is characterized using two complementary metrics: the average number of iterations and the average number of layers involved in message passing per decoded frame. The former captures the convergence behavior of the iterative decoding process, while the latter reflects the effective processing steps during the iterative decoding process. Together, these metrics provide a fine-grained and hardware-relevant characterization of decoding latency.

\subsubsection{Computational Complexity Metrics}
In the proposed decoding algorithm, the computational complexity is primarily dominated by message updates at sum nodes and equality nodes, as well as by APP calculations. In this work, we adopt an operation-level complexity characterization that closely reflects practical decoder implementations. Specifically, the complexity is measured in terms of the average numbers of elementary operations per decoded frame, including addition/subtraction operations, multiplication operations, comparison operations, and look-up table~(LUT) accesses.

The proposed latency and complexity metrics enable a transparent and fair evaluation of sub-normal-graph-based BP decoding and form the basis for the numerical results presented in the following subsection.

\subsection{Numerical Results}
In this subsection, simulation results are presented to demonstrate that the the BP decoding performance of the proposed IBBT polar codes surpasses that of the original BBT polar codes. In addition, the error-rate performance, decoding latency, and computational complexity of the proposed sub-normal-graph-based BP decoding algorithm are evaluated. Two code configurations with $N=300,K=150$ and $N=600,K=300$ are considered.

\begin{figure}[!t]
	\centering
	\includegraphics[width=\linewidth]{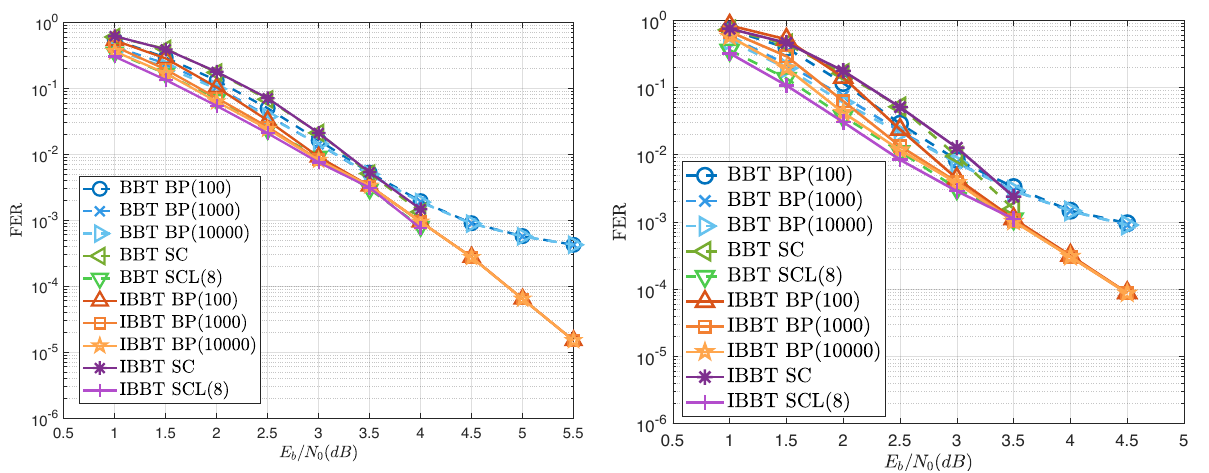}
	\vspace{-2em}
	\caption{FER performance comparison between conventional BBT polar codes and IBBT polar codes under BP decoding for $N=300,K=150$ and $N=600,K=300$.}
	\label{BBTvsIBBT}
\end{figure}
\begin{figure}[!t]
	\centering
	\includegraphics[width=\linewidth]{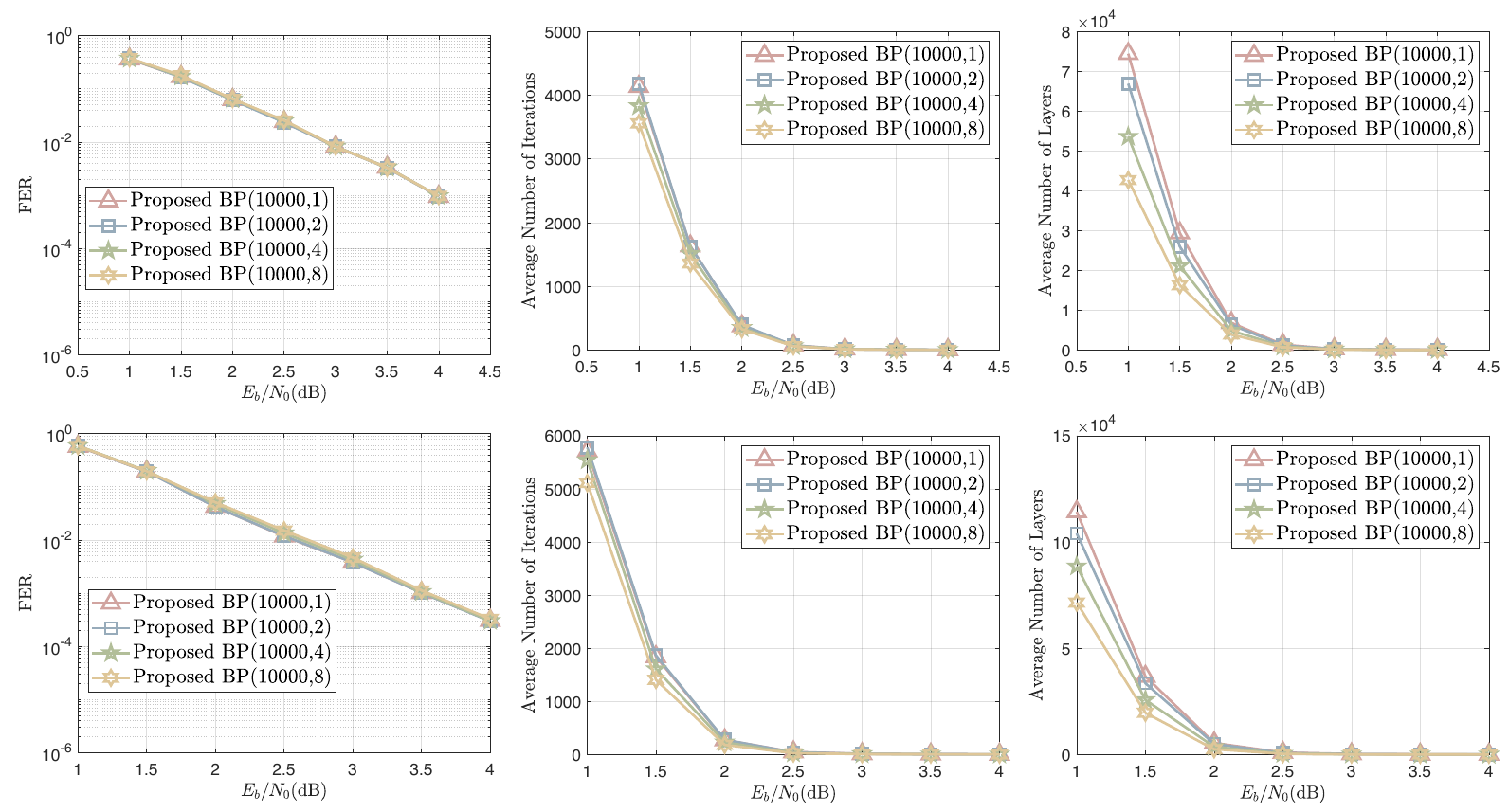}
	\vspace{-3em}
	\caption{Comparison between BP decoding on the full normal graph and the sub-normal graph in terms of error-rate performance and latency-related metrics. Results are shown for $N=300,K=150$ (top row) and $N=600,K=300$ (bottom row). From left to right: FER performance, average number of iterations, and average number of layers involved in message updates.}
	\label{EbN0FERLatency_ProposedBP}
\end{figure}
\begin{figure}[!t]
	\centering
	\includegraphics[width=\linewidth]{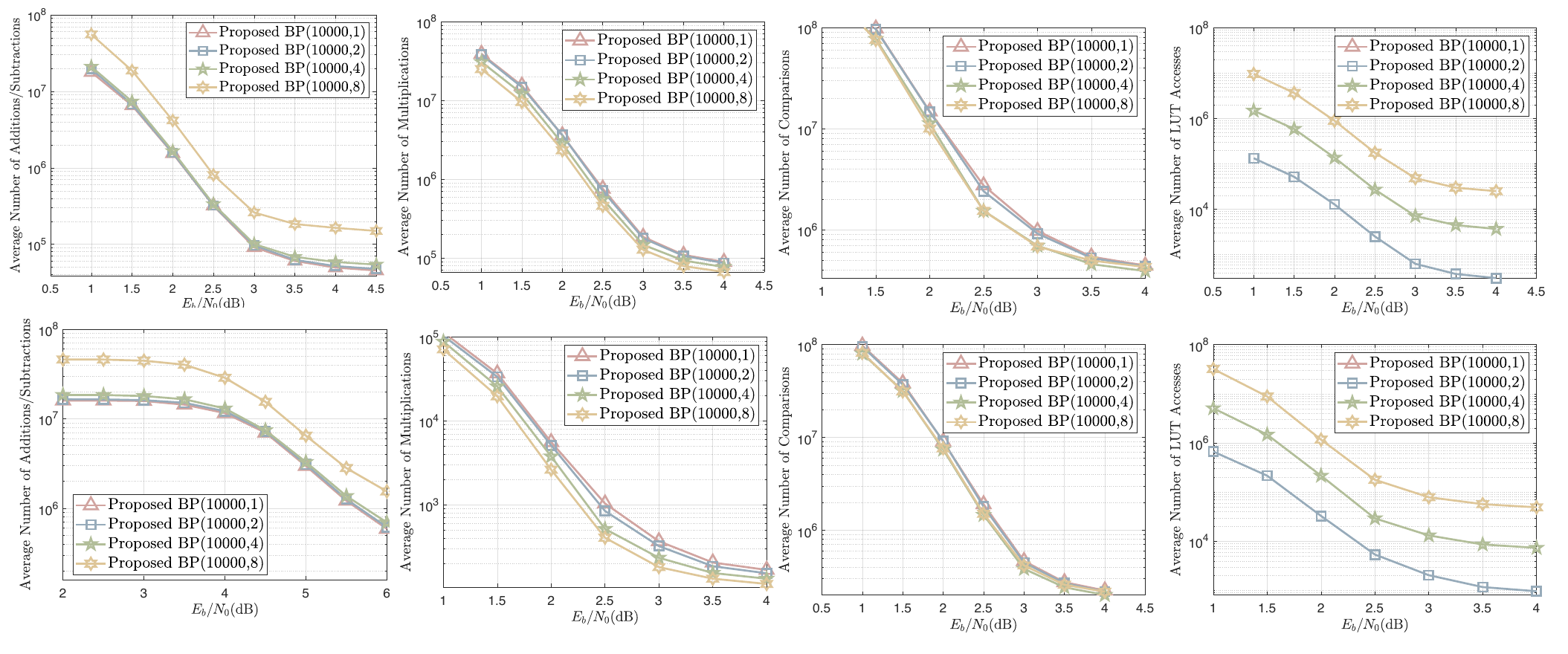}
	\vspace{-3em}
	\caption{Comparison between BP decoding on the full normal graph and the sub-normal graph in terms of computational complexity-related metrics. Results are shown for $N=300,K=150$ (top row) and $N=600,K=300$ (bottom row).}
	\label{EbN0Complexity_ProposedBP}
\end{figure}
Fig.~\ref{BBTvsIBBT} compares the BP decoding performance of BBT polar codes and IBBT polar codes for $N=300,K=150$ and $N=600,K=300$ under different maximum iteration limits, where ${\rm BP}(I_{\rm max})$ denotes BP decoding with a maximum number of iterations $I_{\rm max}$. In both cases, we see that
\begin{itemize}
\item For the original BBT polar codes, increasing $I_{\rm max}$ results in only marginal performance improvement, indicating that BP decoding quickly reaches a saturation point. In contrast, IBBT polar codes continue to benefit from additional iterations, suggesting improved convergence behavior under BP decoding.
\item IBBT polar codes exhibit a significantly lower error floor than BBT polar codes under BP decoding.
\item While BBT and IBBT polar codes have comparable SC and SCL(8) decoding performance, BP decoding applied to IBBT polar codes approaches the performance of SCL(8) decoding, demonstrating its effectiveness.
\end{itemize}

Fig.~\ref{EbN0FERLatency_ProposedBP} compares the error-rate performance and latency-related metrics of our proposed BP decoding algorithm, where ${\rm BP}(I_{\rm max},\ell_{\rm max})$ denotes BP decoding with a maximum of $I_{\rm max}$ iterations and a maximum leaf-subcode length $\ell_{\rm max}$. The results show that BP decoding on the sub-normal graph achieves performance comparable to that on the full normal graph~(i.e., $\ell_{\rm max}=1$) for both block lengths. Meanwhile, both the average number of iterations and the average number of message-passing layers are significantly reduced as $\ell_{\rm max}$ increases, particularly in the low-to-moderate $E_b/N_0$ regime. This effect becomes more pronounced for the longer block length. Moreover, despite a large $I_{\rm max}$, the average number of iterations decreases rapidly with increasing $E_b/N_0$.

Fig.~\ref{EbN0Complexity_ProposedBP} presents the average numbers of elementary operations required for decoding, including additions/subtractions, multiplications, comparisons, and LUT accesses. Compared with the full normal graph, BP decoding on the sub-normal graph significantly reduces the numbers of multiplication and comparison operations for both block lengths, and the reduction is most evident in the low-to-moderate $E_b/N_0$ regime. Although the use of APP calculation modules introduces additional additions and LUT accesses, this overhead remains bounded by the small maximum subcode length $\ell_{\rm max}$.

Overall, the proposed sub-normal-graph-based BP decoding algorithm provides a favorable trade-off: it significantly reduces decoding latency and computational complexity while maintaining performance close to that of conventional BP decoding when $\ell_{\rm max}$ is properly chosen.
\section{Conclusion}
In this paper, we investigated both the theoretical foundations and practical implementation aspects of BBT polar codes. We first proved that the BBT channel transformation induces channel polarization. Consequently, BBT polar codes under SC decoding can achieve the capacity of BMS channels, thereby extending the capacity-achieving property of original polar codes to arbitrary block lengths.

To analyze the finite-length performance of BBT polar codes, we developed an efficient method for estimating the WEF by exploiting the hierarchical tree structure of the BBT construction. Based on the estimated WEF, analytical upper and lower bounds on the FER under ML decoding were derived. Numerical results show that these bounds tightly characterize the ML performance in the high-SNR regime and provide reliable performance prediction in the low-FER region.

For practical low-latency implementations, we further proposed IBBT polar codes together with the sub-normal graph-based BP decoding algorithm. By introducing interleavers between adjacent layers of the normal graph, the convergence behavior of BP decoding can be significantly improved. In addition, BP decoding is performed on an IBBT sub-normal graph, where partial BP processing modules are replaced with APP calculation modules, thereby reducing the number of message-passing steps required per iteration. Numerical results demonstrate that the proposed interleaving strategy improves decoding convergence, while the sub-normal-graph-based BP decoding algorithm significantly reduces decoding latency while maintaining comparable error-rate performance.

Overall, the results of this work establish BBT polar codes as a theoretically sound and practically attractive framework for flexible-length channel coding. Furthermore, the proposed interleaver-assisted BP decoding approach highlights the potential of BBT polar codes for low-latency communication systems.
\section*{Appendix}
\begin{proof}[Proof of Lemma 4]
For a block of $\ell$ independent channels undergoing the BBT polar transformation, we construct a factor graph with $\delta=\lceil \log_2 \ell \rceil$ polarization stages and $\delta+1$ layers, as described in Section II-B. and then associate with each node a subset of $[\delta]\triangleq\{0,\dots,\delta-1\}$ as follows. 

Let $S_{i,j}$ denote the subset associated with the node at position $j$ in layer $i$. At layer $0$, we set $S_{0,j}=\varnothing$ for $0\le j\le \ell-1$. For $1\le i\le\delta$, suppose a polar transformation combines the pair $(j_0,j_1)$ from layer $i-1$ with $j_0<j_1$. We define
\begin{align}\label{LabelRules}
S_{i,j_0}=S_{i-1,j_0}, \qquad
S_{i,j_1}=S_{i-1,j_1}\cup\{i-1\}.
\end{align}
After $\delta$ stages, we obtain a sequence of $\ell$ subsets
\begin{align}
\mathcal{L}_\ell^{(\delta)}=(S_{\delta,0},S_{\delta,1},\dots,S_{\delta,\ell-1}), 
\end{align}
which represents the polarization result of the block.

We first show that the labels, $S_{\delta,j}\in[\delta]$, $0\leq j\leq \ell-1$, are distinct. This follows by induction on the layer index. At layer $0$, all labels are identical but correspond to different channel positions. Each polar transformation replaces one label by itself and the other by the union of that label with a new index that has not previously appeared at that layer. Hence two channels within the same block cannot acquire identical subset labels. Therefore $\mathcal{L}_\ell^{(\delta)}$ consists of $\ell$ distinct subsets of $[\delta]$.

To illustrate the evolution of these subset labels across the layers of the BBT factor graph, we consider the following example.

\begin{example}
We take $\ell=6$ as an example, as illustrated in Fig.~\ref{IllustrationOfBBTPolarization}. 

At layer $0$, all channels can be regarded as forming a single block of size one, and each is associated with the empty set. That is, $\mathcal{L}^{(0)}_{1}=(\emptyset)$. The BBT transformation then proceeds stage by stage. At each stage, pairs of channels are combined according to the factor graph, and the associated subsets are updated according to~(\ref{LabelRules}). Channels belonging to the same newly formed block are marked with the same color in Fig.~\ref{IllustrationOfBBTPolarization} to illustrate how the BBT transformation progressively merges smaller blocks into larger ones while updating the corresponding subset labels.

At layer $1$, three independent polar transformations are performed, producing blocks of size two. Applying~(\ref{LabelRules}) yields $\mathcal{L}^{(1)}_{2}=(\emptyset,\{0\})$.

At layer $2$, some of these blocks are merged to produce a block of size four, yielding $	\mathcal{L}^{(2)}_{4}=(\emptyset,\{1\},\{0\},\{0,1\})$.

At layer $3$, a block of size four and a block of size two are further combined, resulting in $	\mathcal{L}^{(3)}_{6}=(\emptyset,\{2\},\{1\},\{0\},\{0,2\},\{0,1\})$.
\begin{figure}[t]
\centering
\includegraphics[width=0.5\textwidth]{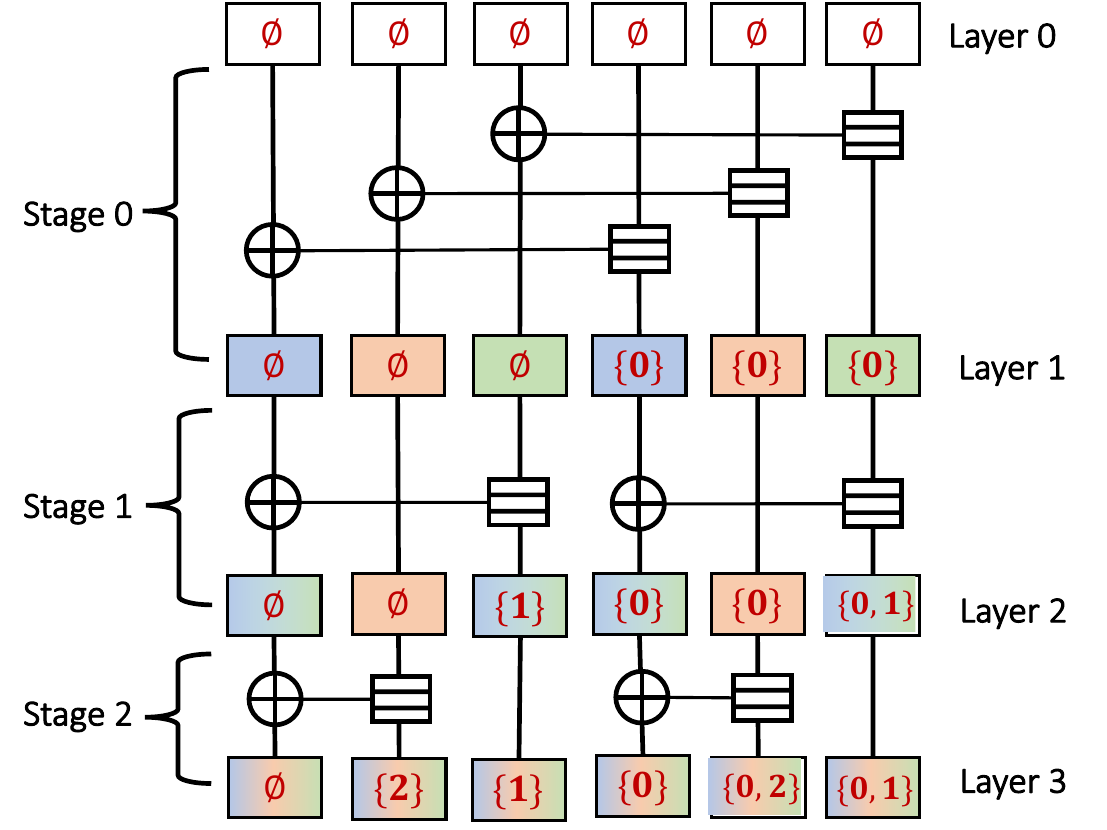}
\caption{Illustration of the BBT polar transformation for $\ell=6$ channels.}
\label{IllustrationOfBBTPolarization}
\end{figure}
\end{example}

Next, recall Definition~1 and denote by $B$, $M$, and $G$ the bad, mediocre, and good channel types, respectively. For a block of $\ell$ channels with polarization result $\mathcal L_\ell^{(\delta)}$, define its pattern by the function
\begin{align}
u:\mathcal L_\ell^{(\delta)}\to\{B,M,G\}.
\end{align}
Equivalently, $u$ may be viewed as the vector $(u(S_{\delta,0}),\dots,u(S_{\delta,\ell-1}))\in\{B,M,G\}^{\ell}$, where $u(S)$ denotes the type of the channel associated with subset $S$.

We claim that if $S,T\in\mathcal L_\ell^{(\delta)}$ satisfy $S\subset T$, then it is impossible that both $u(S)=M$ and $u(T)=M$ when the block is extremal.

To prove this, note that each polar transformation at layer $i$ produces a pair of sets $(S',\,S'\cup\{i\})$. Subsequent operations only append larger indices to both sets simultaneously, thereby preserving this inclusion relation. Hence if $S\subset T$ holds for two sets in $\mathcal L_\ell^{(\delta)}$, they must originate from the same polarization pair $(S',S'\cup\{i\})$ at some layer $i$. Since the block is extremal, the two channels in any polarization pair cannot both be mediocre. Therefore $u(S)$ and $u(T)$ cannot both equal $M$.

Now, define the mediocre-label set
\begin{align}
\mathcal{F}(u)=\{S\in\mathcal L_\ell^{(\delta)}:u(S)=M\}.
\end{align}
From the above argument, no two distinct elements of $\mathcal{F}(u)$ are comparable under set inclusion. Hence $\mathcal{F}(u)$ forms an antichain in the poset $(\mathcal L_\ell^{(\delta)},\subset)$.

Since $\mathcal L_\ell^{(\delta)}\subseteq 2^{[\delta]}$,
any antichain in $\mathcal L_\ell^{(\delta)}$ is also an
antichain in the Boolean lattice $(2^{[\delta]},\subset)$. By Sperner's theorem,
\begin{align}
|\mathcal F(u)|\le\binom{\delta}{\lfloor \delta/2 \rfloor}.
\end{align}
Therefore the number of mediocre channels in the block is at most $\binom{\delta}{\lfloor \delta/2 \rfloor}$.
\end{proof}

\ifCLASSOPTIONcaptionsoff
\newpage
\fi
\bibliographystyle{IEEEtran}
\bibliography{IEEEabrv,RefYaoxym}

\end{document}